\documentclass[a4paper,12pt]{article}
\pdfoutput=1
\usepackage[usenames]{color}
\usepackage[colorlinks=true
,urlcolor=blue
,anchorcolor=blue
,citecolor=green
,filecolor=blue
,linkcolor=blue
,menucolor=blue
,pagecolor=blue
,linktocpage=true
]{hyperref}
\usepackage[centertags]{amsmath}
\usepackage{amsfonts}
\usepackage{amssymb}
\usepackage{bbm}
\usepackage{amsthm}
\usepackage{newlfont}
\usepackage[a4paper]{geometry}
\usepackage{graphicx}
\usepackage{feynmf}

\newcommand{\deltaperms}{
\left(
\delta_{i_1i_2}\delta_{i_3i_4}\delta_{i_5i_6} + 14 \text{\ perms.}
\right)
}

\DeclareMathOperator{\trace}{Tr}

\newcommand{\e}{\epsilon}
\newcommand{\p}{\partial}
\newcommand{\om}{\omega}
\newcommand{\olra}{\overleftrightarrow}
\newcommand{\ora}{\overrightarrow}
\newcommand{\ola}{\overleftarrow}

\newcommand{\ed}{\,.}
\newcommand{\ec}{\,,}
\newcommand{\ecq}{\ec\quad}

\newcommand{\cD}{\ensuremath{\mathcal{D}}}

\newcommand{\cN}{\ensuremath{\mathcal{N}}}
\newcommand{\cO}{\ensuremath{\mathcal{O}}}

\begin{document}


\title{
$d=3$ Bosonic Vector Models Coupled to Chern-Simons Gauge Theories}
\author{Ofer Aharony, Guy Gur-Ari, and Ran Yacoby\\\\
{\it Department of Particle Physics and Astrophysics}\\
{\it Weizmann Institute of Science, Rehovot 76100, Israel}\\
{\small{\tt e-mails~:~Ofer.Aharony,~Guy.GurAri,~Ran.Yacoby@weizmann.ac.il}}}

\maketitle
\vspace{-7mm}
\begin{abstract}
  We study three dimensional $O(N)_k$ and $U(N)_k$ Chern-Simons theories coupled to a scalar field in the fundamental representation, in the large $N$ limit. For infinite $k$ this is just the singlet sector of the $O(N)$ ($U(N)$) vector model, which is conjectured to be dual to Vasiliev's higher spin gravity theory on $AdS_4$. For large $k$ and $N$ we obtain a parity-breaking deformation of this theory, controlled by the 't Hooft coupling $\lambda = 4 \pi N / k$.  For infinite $N$ we argue (and show explicitly at two-loop order) that the theories with finite $\lambda$ are conformally invariant, and also have an exactly marginal $(\phi^2)^3$ deformation.
  For large but finite $N$ and small 't Hooft coupling $\lambda$, we show that there is still a line of fixed points parameterized by the 't Hooft coupling $\lambda$. We show that, at infinite $N$, the interacting non-parity-invariant theory with finite $\lambda$ has the same spectrum of primary operators as the free theory, consisting of an infinite tower of conserved higher-spin  currents and a scalar operator with scaling dimension $\Delta=1$; however, the correlation functions of these operators do depend on $\lambda$. Our results suggest that there should exist a family of higher spin gravity theories, parameterized by $\lambda$, and continuously connected to Vasiliev's theory. For finite $N$ the higher spin currents are not conserved.
\end{abstract}

\newpage

\tableofcontents

\setlength{\unitlength}{1mm}

\section{Introduction}

The $AdS$/CFT correspondence \cite{Maldacena:1997re} is an exact duality between quantum gravitational theories on
space-times that include anti-de Sitter space $AdS_{d+1}$, and conformal field theories in $d$ space-time dimensions. This correspondence has many applications, and it has taught us a lot about strongly coupled field theories and about quantum gravity. However, while we know how to translate computations on one side of the duality to the other side, we do not yet have a derivation of the $AdS$/CFT correspondence, that would enable us in particular to know which quantum gravity theory is dual to a given conformal field theory, and vice versa. Finding such a derivation is complicated by the fact that in most examples, either one or both sides of the correspondence are strongly coupled. This is partly because the gravitational dual of any weakly coupled field theory must include light fields of arbitrarily high spin.

There is one example of the $AdS$/CFT correspondence in which both sides are  weakly coupled in the large $N$ limit; this is the conjectured duality \cite{Sundborg:2000wp,Sezgin:2002rt,Klebanov:2002ja} between the singlet sector of the $O(N)$ vector model (namely, $N$ free real scalar fields) in three space-time dimensions, and Vasiliev's higher-spin gravity theory on $AdS_4$ \cite{Fradkin:1987ks} (see \cite{Vasiliev:1999ba} for a review). While the gravitational side of this duality is only understood at the classical level, and it is not yet known how to give it a quantum completion, in the classical gravity limit (governed by tree-level diagrams in the bulk) this provides an example of the $AdS$/CFT correspondence in which both sides are weakly coupled. This allows many detailed tests of the correspondence to be performed in this case \cite{Giombi:2009wh,Giombi:2010vg,Giombi:2011ya}, and it also suggests that this could be an ideal toy model for which a derivation of the $AdS$/CFT correspondence could be found (and perhaps then generalized to more complicated cases). Indeed, there are several suggestions in the literature \cite{Das:2003vw,Koch:2010cy,Douglas:2010rc,Jevicki:2011ss} for how to derive the $AdS$/CFT correspondence explicitly in this example.

In this paper we study a small deformation of the duality above, on the field theory side; it should be possible to map any such deformation to the gravity side as well, and to utilize the extra structure that it provides to learn more about the explicit $AdS$/CFT mapping in this case. A simple way to obtain a theory that contains only the singlet sector of the $O(N)$ vector model is by coupling $N$ free scalar fields to an $O(N)$ gauge theory; since we do not want to add any dynamics of the gauge field, we should not have standard kinetic terms for the gauge fields, but we can view their action as the $k\to \infty$ limit of the $O(N)_k$ Chern-Simons gauge theory \cite{Giombi:2009wh}.\footnote{The Chern-Simons action is required to make the operator $F_{\mu \nu}(x)$ trivial, to ensure that we do not add any additional local operators to the theory beyond the singlets of the vector model.} It is then natural to deform the theory by making the Chern-Simons level $k$ a finite integer; this theory has a 't Hooft limit, controlled by a 't Hooft coupling $\lambda \equiv 4 \pi N / k$, and at large $N$ this gives a continuous parity-breaking deformation of the original theory. On the field theory side one can then perform computations in perturbation theory in $\lambda$, and it should be possible to translate these into perturbative computations also on the gravity side, and to obtain a more detailed weak-weak coupling duality. We will consider both the $O(N)$ case with a real scalar field in the fundamental representation, and the $U(N)$ case with a complex scalar. These theories were previously studied perturbatively in \cite{Chen:1992ee,Avdeev:1992jt}.

We begin in section \ref{On} by introducing our action and our methods of regularizing and renormalizing it. In section \ref{confSym} we study whether the theory at small $\lambda$ is still conformally invariant. The Chern-Simons level is quantized and does not run \cite{Deser:1981wh,Deser:1982vy}. One problem that may arise whenever we have interactions is that relevant operators of the form $\phi^2$ and $(\phi^2)^2$ (where $\phi$ is the scalar field) may be generated, even if they are tuned to zero at some scale. However, in our renormalization scheme these couplings do not run away if they are initially set to zero. A more serious problem is that this theory has a classically marginal $\lambda_6 (\phi^2)^3 / N^2$ coupling, which could start running once we turn on $\lambda$. However, we provide an argument (and check explicitly at two-loop order) that at infinite $N$ the beta function for this coupling vanishes. Therefore, there is a two-dimensional family of large $N$ conformal field theories, parameterized by $\lambda$ and by $\lambda_6$.
For large but finite $N$ we show that a beta function for $\lambda_6$ is generated, but that (at least) for small $\lambda$ this beta function still has an IR-stable fixed point, so that there still exists a one-parameter family of conformal field theories, parameterized by $\lambda$. Note that while $\lambda$ is a discrete parameter for finite $N$, it is almost continuous when $N$ is very large.

In section \ref{HScurrents} we analyze the spectrum of the large $N$ family of conformal field theories that we find, and show that it is independent of $\lambda$ (and thus identical to that of the free theory with $\lambda=0$). In particular, conserved higher-spin currents still exist for infinite $N$ and any $\lambda$, though the corresponding symmetries are broken for finite $N$. Such an appearance of an infinite number of conserved currents in an interacting theory is quite surprising, and this could lead one to suspect that the theories we discuss may be independent of $\lambda$ in the large $N$ limit. In section \ref{conjecture} we show that this is not the case, by computing a correlation function in these theories at leading order in $\lambda$ (in the large $N$ limit) and showing that it depends on $\lambda$. We end in section \ref{summary} with a summary of our results and a discussion of some future directions.

\section{The $O(N)$ Model with Chern-Simons Interactions}
\label{On}

Consider the theory of a real scalar field $\phi$ in the fundamental representation of $O(N)$, coupled to gauge bosons $A_\mu$ with Chern-Simons interactions at level $k$ in three Euclidean dimensions (the generalization to $N$ complex scalar fields coupled to a $U(N)_k$ Chern-Simons theory is straightforward, and we will occasionally discuss below this case as well). We regulate the theory using dimensional reduction \cite{Siegel:1979wq} (see below), and work in Lorenz gauge (Landau gauge), $\partial^{\mu}A_\mu=~0$. The regularized action in terms of the renormalized fields and couplings is
\begin{align}
  S &= S_\mathrm{CS} + S_\mathrm{gh} + S_\mathrm{b}\ec \label{eq:action}\\
  S_\mathrm{CS} &= \int \! d^dx\, \left\{
  - \frac{i}{2} Z_A
  \epsilon_{\mu\nu\lambda} A_\mu^a \partial_\nu A_\lambda^a
  - \frac{i}{6} \mu^{\epsilon/2} g Z_g \epsilon_{\mu\nu\lambda} f^{abc}
  A_\mu^a A_\nu^b A_\lambda^c
  \right\}\ec \label{eq:actionCS}\\
  S_\mathrm{gh} &= \int \! d^dx\, \left\{
  - \frac{1}{2 \gamma_R} (\partial_\mu A_\mu^a)^2
  + Z_\mathrm{gh} \partial_\mu \bar{c}^a \partial^\mu c^a
  + \mu^{\epsilon/2} \tilde{Z}_g g f^{abc}
  \partial_\mu \bar{c}^a A_\mu^b c^c
  \right\}\ec \label{eq:actionGH}\\
  S_\mathrm{b} &= \int \! d^dx\, \left\{
  \frac{1}{2} Z_\phi (\partial_\mu \phi_i)^2
  + \mu^{\epsilon/2} Z_g' g \partial_\mu \phi_i T_{ij}^a
  A_\mu^a \phi_j
  - \frac{1}{4} \mu^\epsilon Z_g'' g^2 \{T^a,T^b\}_{ij}
    \phi_i \phi_j A_\mu^a A_\mu^b
  \right. \notag\\
  &\quad \qquad \qquad \left.
  + \mu^{2\epsilon} Z_{g_6}\frac{g_6}{3!\cdot 2^3}(\phi_i\phi_i)^3
  \right\} \ec
  \label{eq:actionB}
\end{align}
where $d=3-\epsilon$, and $\mu$ is the renormalization scale (for additional conventions, see Appendix \ref{conventions}). The coupling $g$ is related to the integer Chern-Simons level $k$ by $k = 4\pi/g^2$. When taking the 't Hooft large $N$ limit, the couplings $\lambda = g^2 N$ and $\lambda_6 = g_6 N^2$ are held fixed, and in this limit $\lambda$ becomes a continuous parameter. Note that while parity is broken due to the Chern-Simons interaction, the theory is dual under the combined transformation of parity plus $\lambda\to-\lambda$ ($k\to -k$), and physical results must be invariant under this transformation.

Once $\lambda > 0$, in order to renormalize the theory in a generic scheme we must add also two relevant interactions that will be generated by quantum corrections: a mass term $(\phi_i \phi_i)$ and an interaction of the form $g_4 (\phi_i \phi_i)^2$. We are interested in interacting conformal fixed points of our field theory, so we will generally fine-tune our couplings so that the physical mass and $\phi^4$ couplings vanish,
and then for the purposes of our computations we can just ignore these terms. In fact, in the
scheme we are using (of dimensional reduction and minimal subtraction), once we fix the
renormalized dimensionful couplings to zero, they remain zero so we do not even have to add them to our action.

At least in the large $N$ limit, we could also study the theory in which the coupling $g_4$ does not vanish;
if it is non-zero then the theory flows to another fixed point, which at large $N$ is closely related to the original fixed point (at infinite $N$ it has the same spectrum of operators, except for the operator $\phi_i \phi_i$ whose dimension at the interacting fixed point is $\Delta=2$). For the theory with $\lambda=0$ this was discussed in the $AdS$/CFT context in \cite{Klebanov:2002ja,Gubser:2002vv,Petkou:2003zz,Giombi:2011ya}, and the same analysis holds also at finite $\lambda$. Therefore, most of our results also apply to the ``critical'' fixed point with a non-zero $g_4$ coupling. However, for simplicity, we will focus here on the case where the physical $g_4$ coupling is tuned to vanish.

Note that dimensional regularization of this theory is subtle, since the 3-form integration of the Chern-Simons interaction (\ref{eq:actionCS}) is not well-defined for arbitrary dimension. To regulate loop integrals we first perform the tensor algebra in 3 dimensions, and then compute the resulting scalar integral in $d=3-\epsilon$ dimensions. This method, known as dimensional reduction \cite{Siegel:1979wq}, has been shown in \cite{Chen:1992ee} to preserve gauge-invariance in this theory at least up to two-loop order.

\section{Conformal Symmetry}
\label{confSym}

In this section we analyze the conditions under which the theory defined by \eqref{eq:action} is conformal, both for finite and for infinite $N$. The Chern-Simons level $k$ is quantized to be an integer and is therefore not renormalized, except perhaps by an integer shift at one-loop order; this has been verified explicitly in \cite{Chen:1992ee}. The corresponding one-loop shift in $\lambda$ is of order $1/N$ in the 't Hooft large-$N$ limit that we study here, so we will ignore it. However, the classically-marginal $\lambda_6$ coupling may receive corrections.\footnote{For the Abelian Chern-Simons theory coupled to a scalar field, such corrections were studied in \cite{Alves:1999hw,deAlbuquerque:2000ec}.} In order to check for conformal fixed points we need to compute its beta function $\beta_{\lambda_6}(\lambda,\lambda_6)$, and show that it vanishes. In section \ref{2loop} we compute this beta function at the first non-trivial order, by computing the divergent contributions to the amputated correlator
\begin{align}\label{sixpoint}
  \left< \phi^{i_1}(x_1) \cdots \phi^{i_6}(x_6) \right>_{\mathrm{amp.}} \ed
\end{align}
We might expect that solving $\beta_{\lambda_6}(\lambda,\lambda_6)=0$ would result in a line of fixed points in the $(\lambda,\lambda_6)$ plane. For large and finite $N$ we indeed find two such lines; however, at infinite $N$ we find that $\lambda$ and $\lambda_6$ are both exactly marginal at 2-loops. In section \ref{allorders} we show that this is actually true to all orders in perturbation theory, so that at infinite $N$ there is a family of conformal field theories labeled by continuous parameters $\lambda$ and $\lambda_6$. In section \ref{sponbreak} we argue that there is no spontaneous breaking of the conformal symmetry in our theories.

\subsection{The Beta Function $\beta_{\lambda_6}$ at Two Loops}
\label{2loop}

In this section we compute $\beta_{\lambda_6}$ in momentum space using minimal subtraction. In our theory, using our dimensional reduction regularization, all 1-loop integrals are finite. Indeed, for quadratic and logarithmic divergences in three dimensions, the numerator must be an odd power of the loop momentum $q$, which must be of the form $q^\mu q^{2n}$, and then the $q$ integral vanishes by the $q\to -q$ symmetry. Linear divergences are rendered finite by dimensional regularization. In the specific case of $\beta_{\lambda_6}$, a one-loop contribution is also not allowed by the parity transformation.

Therefore, the leading contribution to this beta function arises at 2-loop order. The $\left<\phi^6\right>$ correlator (\ref{sixpoint}) is superficially log-divergent, with over 50 two-loop diagrams contributing to it in the planar limit alone. However, the number of diagrams that may contribute to its divergence is greatly reduced by the following observation. Consider a diagram that includes a $\phi A^\mu \partial_\mu \phi$ vertex, with the gluon carrying a loop momentum $q$ and one of the scalar lines carrying an external momentum $p$. In the numerator we then have $\epsilon_{\mu\nu\rho} q^\rho$ from the gluon propagator (\ref{eq:GluonProp}) and $(q+2p)^\mu$ from the vertex, and the leading high-energy term of order $q^2$ cancels by antisymmetry. Therefore, in such a situation the degree of divergence is reduced and the diagram is finite. As a result, the only diagrams that can contribute to $\beta_{\lambda_6}$ at 2-loop order are the following:

\begin{center}
\begin{tabular}{cccc}
\includegraphics[height=3.5cm]{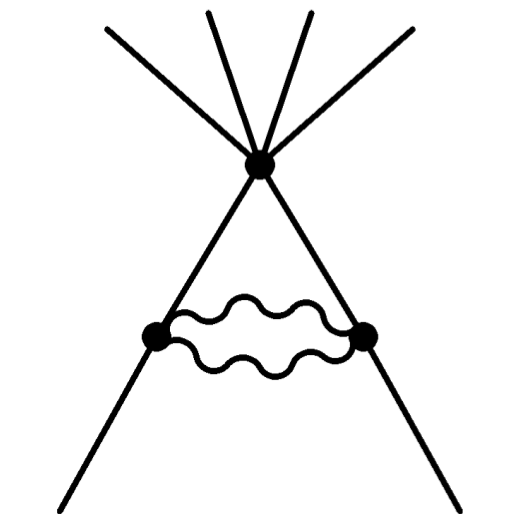} &
\includegraphics[height=3.5cm]{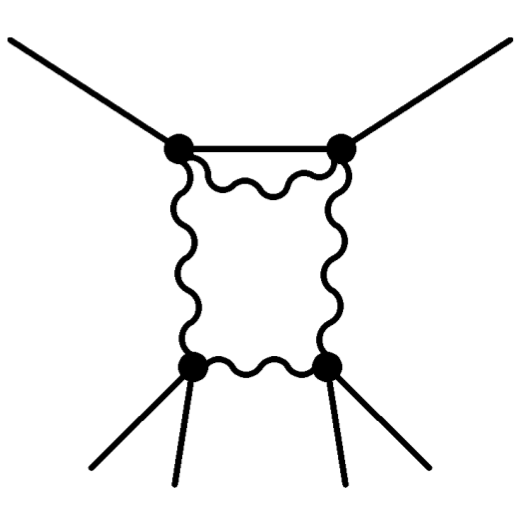} &
\includegraphics[height=3.5cm]{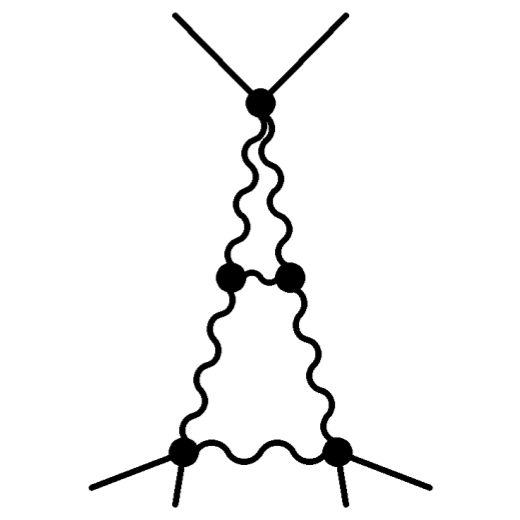} &
\includegraphics[height=3.5cm]{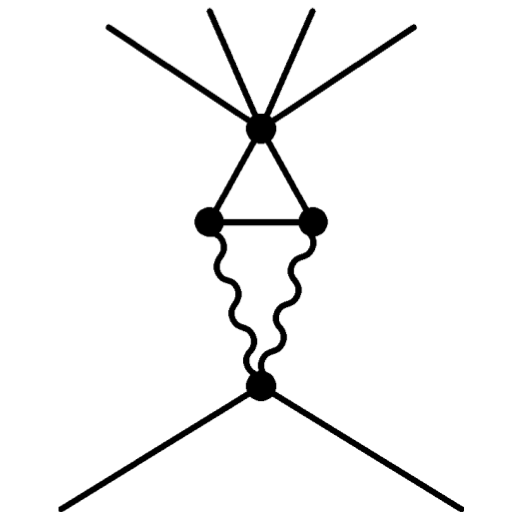}
\\
(A1) & (A2) & (A3) & (A4)
\vspace{8mm}
\\
\includegraphics[height=3.5cm]{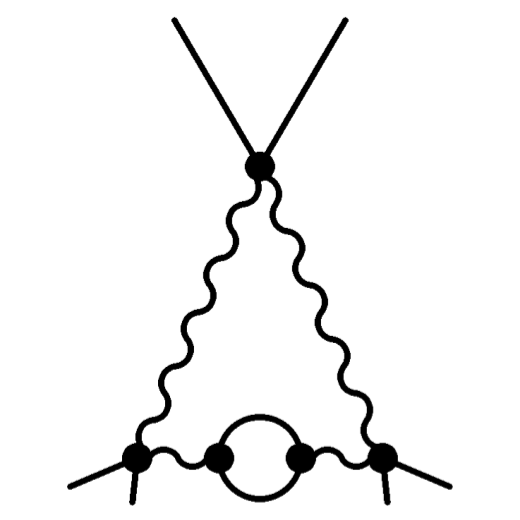} &
\includegraphics[height=3.5cm]{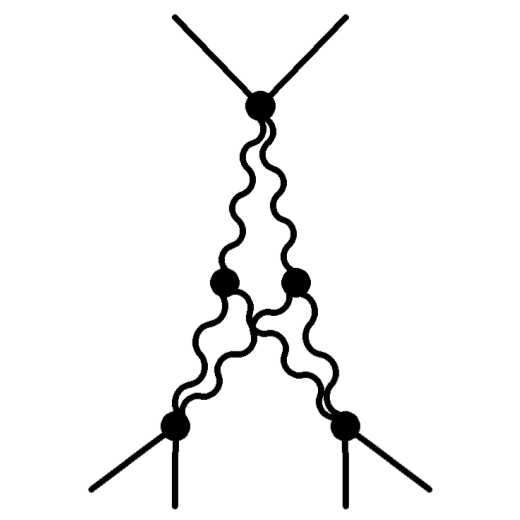} &
\includegraphics[height=3.5cm]{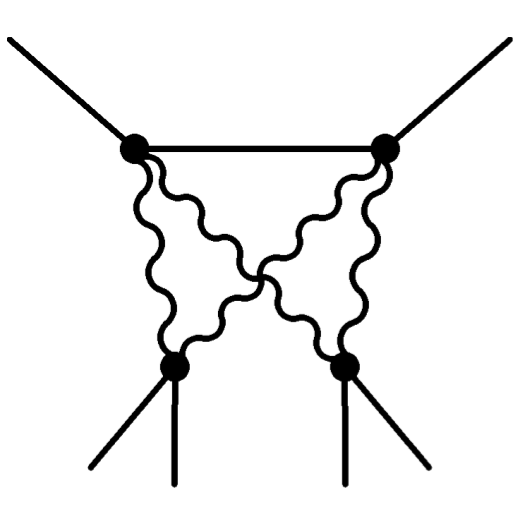} &
\includegraphics[height=3.5cm]{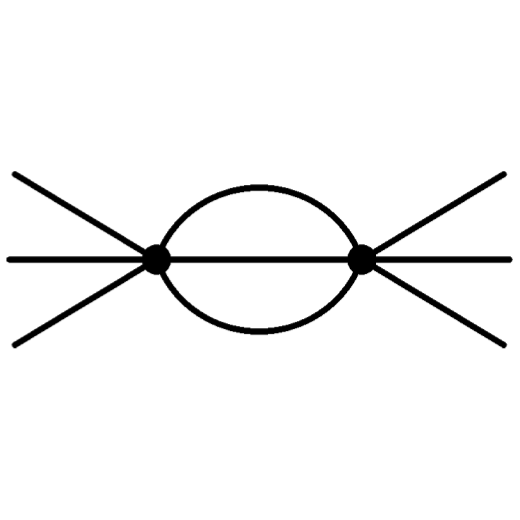}
\\
(A5) & (A6) & (A7) & (A8)
\end{tabular}
\end{center}

The diagrams (A1-3) include planar diagrams, while the others are suppressed by powers of $1/N$ in the 't Hooft large $N$ limit. In order to compute the 2-loop beta function, we need in addition to the diagrams above also the anomalous dimension of the scalar field at this order. This comes from the following diagrams:\footnote{These diagrams were already computed in \cite{Chen:1992ee}.}

\vspace{-5mm}

\begin{center}
\begin{tabular}{cccc}
\includegraphics[width=3.7cm]{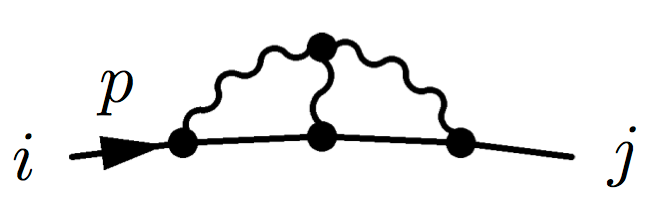} &
\includegraphics[width=3.7cm]{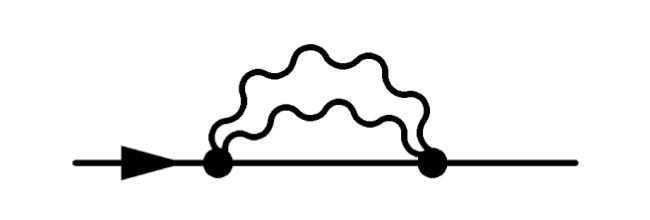} &
\includegraphics[width=3.7cm]{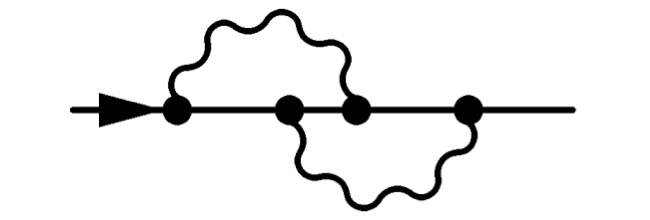} &
\includegraphics[width=3.7cm]{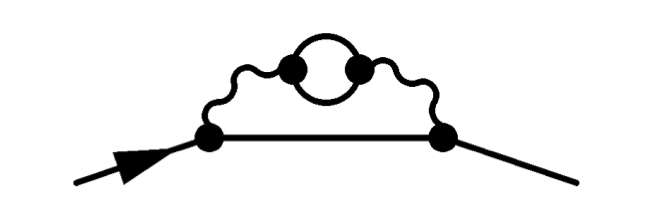}
\\
(B1) & (B2) & (B3) & (B4)
\end{tabular}
\end{center}

The divergent parts of all the diagrams above are listed in Appendix \ref{diags}. By summing these we can determine the renormalization constants,
\begin{align}
  Z_{\phi} &\equiv 1 + \sum_i(Bi) = 1 - \frac{g^4\left(3N^2-23N+20\right)}{384\pi^2}\frac{1}{\epsilon}\ec \label{eq:Zphi}\\
  g_6Z_{g_6} &\equiv g_6 + \sum_i(Ai) = g_6 + \frac{66g^8\left(N-1\right) + 4g_6^2\left(3N+22\right) - 3g^4g_6\left(N^2+19N-20\right)}{128\pi^2}\frac{1}{\epsilon}\ed \label{eq:Zg6}
\end{align}
The bare sextic coupling $g_{6,0} = \mu^{2\epsilon} g_6 Z_{g_6} / Z_{\phi}^3$ may thus be written in the form $g_{6,0} = g_6 + b_1(g,g_6) / \epsilon + (\mathrm{other~terms})$, where
\begin{align}
b_1(g,g_6) = \frac{33 g^8(N-1) - 40 g^4 g_6 (N-1) + 2 g_6^2 (3 N + 22)}{64 \pi^2}\ed
\end{align}
The beta function for the $\lambda_6$ coupling is related to the single pole in dimensional regularization by $\beta_{g_6} = - 2 b_1 + 2 g_6 \partial_{g_6} b_1 + {1\over 2} g \partial_g b_1$ \cite{Weinberg:1996kr}, leading to
\begin{align}\label{eq:Full2LoopBeta6}
\beta_{\lambda_6}(\lambda,\lambda_6) = \frac{33 (N-1) \lambda ^4-40 (N-1) \lambda ^2\lambda_6+2 (3 N + 22) \lambda_6^2}{32 N^2 \pi ^2} \ed
\end{align}

In the 't Hooft large $N$ limit we see that $\beta_{\lambda_6}=0$, so that both the $(\phi^2)^3$ coupling and the Chern-Simons interaction are marginal at this order. For the theory with only $(\phi^2)^3$ couplings it is easy to see that the large $N$ beta function vanishes to all orders, since there are no contributing diagrams; it is indeed well-known that this coupling is exactly marginal in the large $N$ limit \cite{Bardeen:1983rv} (see \cite{Moshe:2003xn} for a review). However, for finite $\lambda$ there do exist divergent planar diagrams. The vanishing of the $\lambda^4$ term in \eqref{eq:Full2LoopBeta6} at this order is due (in our gauge choice) to a non-trivial cancelation between the diagrams (A2) and (A3). There is also a large $N$ divergence proportional to $\lambda^2 \lambda_6$ arising from (A1), that exactly cancels in the planar limit with the similar contribution from the anomalous dimension of $\phi^i$. In fact, one can show that, at large $N$, contributions to the beta function can have either zero or one $(\phi^2)^3$ vertices, and that the planar diagrams contributing with a single $(\phi^2)^3$ vertex are the same as the diagrams contributing to the anomalous dimension of $\phi^2$. Thus, the large $N$ beta function takes the form
\begin{equation}
\beta_{\lambda_6}(\lambda,\lambda_6) = b\, \gamma_{\phi^2}(\lambda)\lambda_6 + f(\lambda) + O(1/N),
\end{equation}
where $\gamma_{\phi^2}$ is the anomalous dimension of $\phi^2$ and $b$ is a constant. In the next subsection we argue that both this anomalous dimension and the beta function $\beta_{\lambda_6}$ vanish in the large $N$ limit, so that the couplings $\lambda$ and $\lambda_6$ are both exactly marginal in this limit.

At finite but large $N$ the beta function does not vanish. Without the coupling $\lambda$, the beta function is positive so the theory with $\lambda_6 > 0$ is trivial (IR-free). However, when $\lambda \neq 0$ and for large $N \geq 10$, we find from \eqref{eq:Full2LoopBeta6} two lines of non-trivial fixed points of the two-loop beta function,
\begin{align}
  \lambda_{6}^\pm(\lambda) &=
  \frac{\left(20 N - 20 \pm \sqrt{1852-2054 N+202 N^2} \right) \lambda^2}
  {44+6 N} \ed
\end{align}
The line $\lambda_6^+(\lambda)$ is IR-stable, while $\lambda_6^-(\lambda)$ is UV-stable -- see Figure 1.
Note that since $\beta_\lambda=0$, the renormalization group flow is always in the $\lambda_6$ direction.

\begin{figure}[!ht]\label{fig:RG}
\centering
  \includegraphics[width=0.6\textwidth]{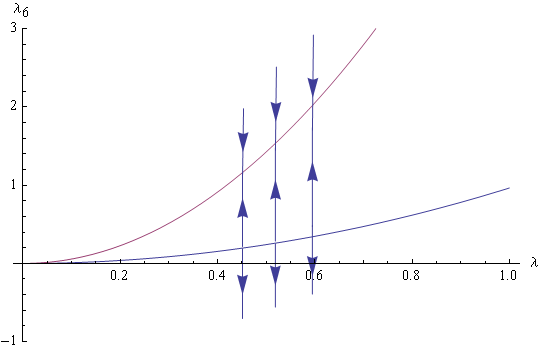}
  \caption{The space of coupling constants and the renormalization group flow towards the IR for large but finite $N$, based on the two-loop beta function. There are two lines of fixed points. Since $\beta_{\lambda}=0$, the flow lines are all in the $\lambda_6$ direction.}
\end{figure}

\subsection{The Large $N$ Beta Function $\beta_{\lambda_6}$ to All Orders}
\label{allorders}

In this section we argue that $\beta_{\lambda_6}(\lambda,\lambda_6) = O(1/N)$ to all orders in perturbation theory, generalizing our explicit two-loop computation of the previous subsection. We could not find a direct argument for this, so instead we will use a trick. We focus on the $U(N)$ vector model for simplicity, but the argument can be generalized to $O(N)$ as well.

Consider the $\cN=2$ supersymmetric generalization of our theory, which is the ${\cN=2}$ supersymmetric Chern-Simons $U(N)$ gauge theory, coupled to a single matter chiral superfield $\Phi_i$ with components $(\phi_i,\psi_i)$ in the fundamental representation (we use $i,j = 1,\ldots,N$ to label the fundamental representation of $U(N)$).\footnote{This theory has a ``parity anomaly'' which means that $k$ must be a half-integer rather than an integer, but this will not affect our large $N$ discussion here.} We will relate $\beta_{\lambda_6}$ in our theory (for infinite $N$) to the beta function of the $(\phi^\dagger\phi)^3$ coupling in the $\cN=2$ theory. The action of the $\cN=2$ theory \cite{Schwarz:2004yj,Gaiotto:2007qi}, after integrating out all the auxiliary fields, is
\begin{align}\label{eq:SUSicCS}
S^{\cN=2}_{CS} &= -\frac{ik}{4\pi}\int \trace\left[A\wedge dA + \frac{2}{3}A^3\right] + \notag\\
&\quad \int \! d^3x \, \left[ |\cD_{\mu}\phi_i|^2 +i\bar{\psi}^i\gamma^{\mu}\cD_{\mu}\psi_i 
- 2\frac{\lambda}{N}\bar{\phi}^i\phi_i\bar{\psi}^j\psi_j - \frac{\lambda}{N}\bar{\phi}^i\phi_j\bar{\psi}^j\psi_i - \frac{\lambda^2}{N^2}(\bar{\phi}^i\phi_i)^3 \right]
\ed
\end{align}
It was shown in \cite{Gaiotto:2007qi} that this action is exactly conformal quantum mechanically, for all values of $k$ and $N$ (with $\lambda=4\pi N/k$). In particular, this means that the beta function of the $(\phi^\dagger\phi)^3$ coupling in this theory vanishes identically to all orders in $\lambda$ and $1/N$. The theory has a global $U(1)_f$ symmetry acting on the matter superfield as $\Phi \to e^{i\alpha}\Phi$, and in \cite{Gaiotto:2007qi} it was noticed that the operators $\cO_1 = \bar{\phi}^i\phi_i$ and $\cO_2 = \bar{\psi}^i\psi_i + \frac{4\pi}{k}(\bar{\phi}^i\phi_i)^2$ sit in the same supermultiplet as the $U(1)_f$ symmetry current.\footnote{The $U(1)$ flavor symmetry is in fact part of the gauged $U(N)$ symmetry. Nevertheless, in the large $N$ limit that we are interested in, we can obtain the same results by gauging only an $SU(N)$ group, and then the $U(1)$ is a global symmetry.} As a consequence, the dimensions of $\cO_1$ and $\cO_2$ are protected to be 1 and 2 respectively. The ``double-trace'' term in $\cO_2$ does not contribute to the 2-point function $\langle\cO_2(x)\cO_2(y)\rangle$ to leading order in $1/N$, and therefore the  operator $\bar{\psi}^i\psi_i$ by itself is also protected at large $N$, with dimension $\Delta=2+O(1/N)$.

Let us begin by arguing that the anomalous dimension of $\cO_1$ vanishes at large $N$ also in our non-supersymmetric theory. Consider the diagrams that contribute to $\left< \cO_1 \cO_1 \right>$ and $\left< \cO_2 \cO_2 \right>$ in the $\cN=2$ theory and involve a single matter loop, with possible additional gluon lines. We will denote them collectively as
\begin{center}
\begin{tabular}{cc}
\includegraphics[height=3.5cm]{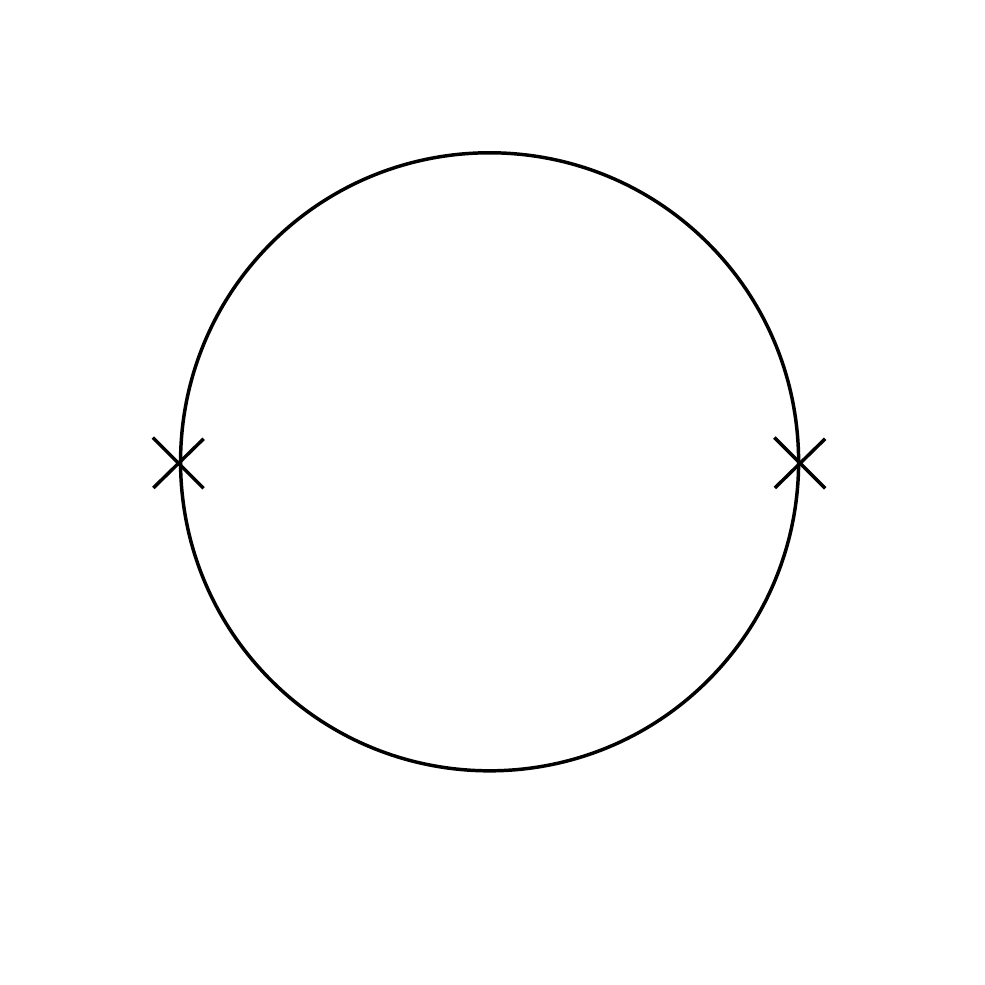} &
\vspace{-2mm}
\includegraphics[height=3.5cm]{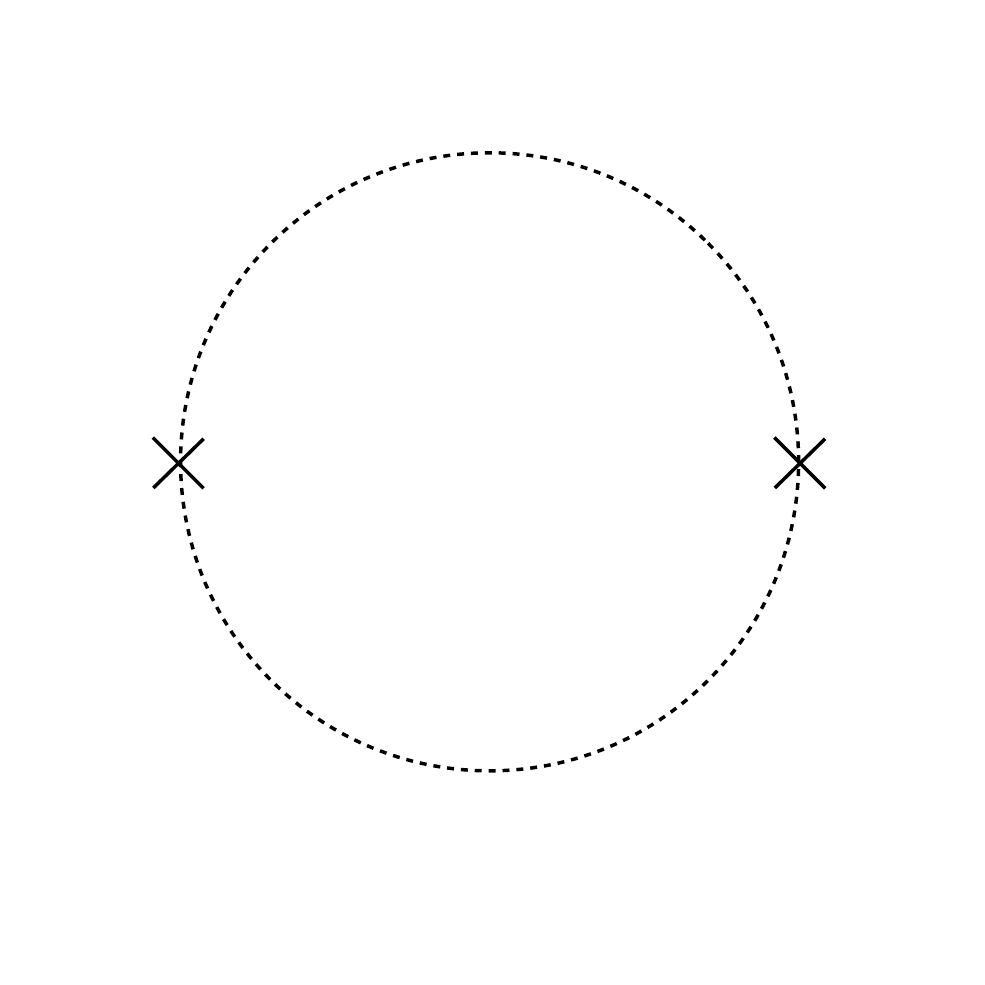}
\vspace{-2mm}
\\
(E1) & (E2)
\end{tabular}
\end{center}
For the rest of the section we will keep gluon lines implicit in all diagrams; at large $N$ when we draw the diagrams in double-line notation these lines must sit inside the scalar/fermion loop so that the topology of each scalar/fermion loop is that of a disk. We will show below that in the large $N$ limit the sum of such diagrams at a given order in perturbation theory is finite. However, the diagrams (E1) (with gluon lines running in the loop) are precisely those that contribute to the correlator $\left< \cO_1 \cO_1 \right>$ in our $\cN=0$ model in the large $N$ limit.\footnote{There are also diagrams that include $(\phi^\dagger\phi)^3$ vertices, but they have tadpole matter loops, and all such loops vanish in our regularization scheme.}
Thus, it will follow that the dimension of $\cO_1 = \bar{\phi}^i \phi_i$ in our non-supersymmetric vector model is $1+O(1/N)$ to all orders in planar perturbation theory.

We now prove the finiteness of (E1) and (E2) at large $N$ by induction. At zeroth order in perturbation theory, (E1) and (E2) are single 1-loop diagrams which are finite in our regularization scheme. At the next order the only diagrams contributing to the two-point functions in the $\cN=2$ theory are still of the form (E1) and (E2) (with an extra gluon line), so all divergences in these diagrams must cancel (in fact, it follows from the parity transformation that these diagrams vanish).
At higher orders in perturbation theory, there are more general diagrams contributing to $\left< \cO_1 \cO_1 \right>$ at large $N$, which have the general form:
\begin{center}
\begin{tabular}{c}
\includegraphics[height=3.5cm]{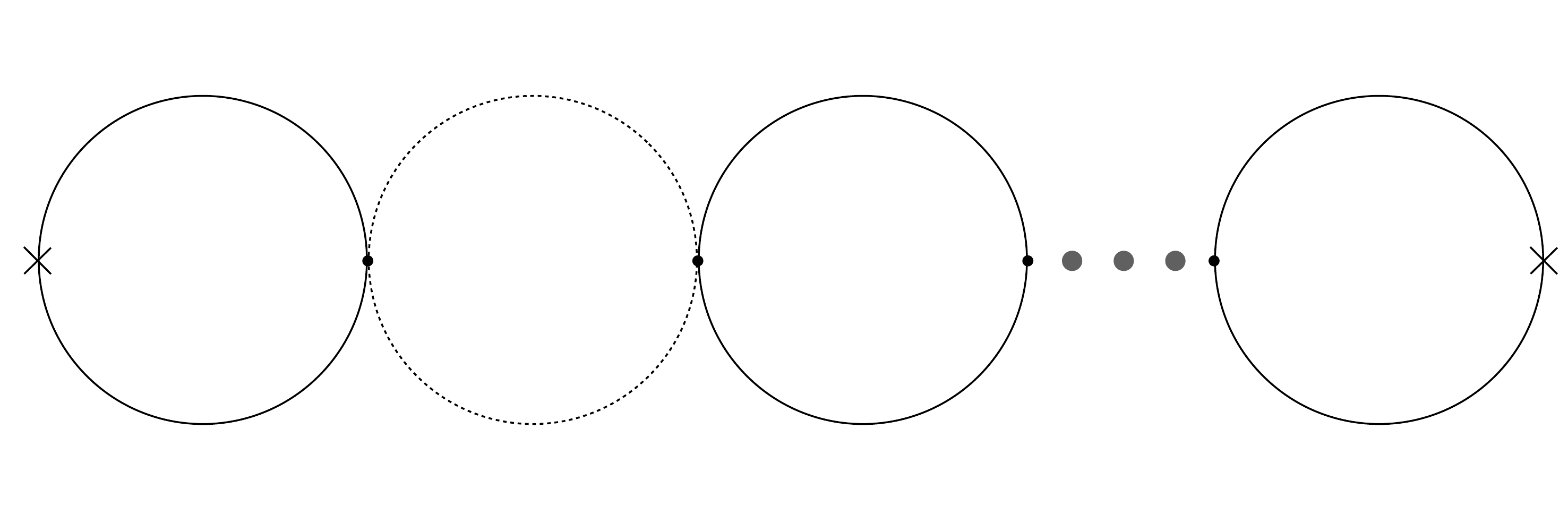}
\vspace{-2mm}
\\
(F1)
\end{tabular}
\end{center}
Again, gluon lines running inside the loops are implicit, and all the other diagrams (not drawn in (F1)) contain tadpole matter loops causing them to vanish.
We know that in the $\cN=2$ theory the sum of all these (F1) diagrams, with any (odd) number of matter loops, is finite, since $\cO_1$ is not renormalized. Working in momentum space, each (F1) diagram factorizes at large $N$ into a product of sub-diagrams of the form (E1) or (E2). If a given (F1) diagram has more than one matter loop, its sub-diagrams will be of a lower order in perturbation theory. The sum over such sub-diagrams is finite by the induction assumption, and therefore (F1) diagrams with more than one matter loop are finite in total. Since the sum over all (F1) diagrams is also finite, the sum over single-matter-loop diagrams~--- which are the (E1) diagrams at the order we are in --- must be finite. This concludes the induction step for (E1); the step for (E2) is analogous.\footnote{Note that we are using here the fact that we only have marginal couplings. In a theory with relevant operators like $(\phi^2)^2$, anomalous dimensions can arise even from finite diagrams, but this is not true in our case.} In appendix \ref{adim} we verify that indeed the anomalous dimension of $\phi^2$ vanishes in the non-supersymmetric theory at two-loop order in the large $N$ limit.

The argument above can be easily generalized to diagrams of the topology (E1), which have three insertions of $\cO_1$ on the scalar loop instead of two. Namely, the sum of such diagrams is also finite (in the large $N$ limit) in the non-supersymmetric theory at a given order in perturbation theory. To see this consider the correlator $\left< (\cO_1)^3 \right>$ in the $\cN=2$ theory, which does not contain divergences since both $\cO_1$ and the $(\phi^\dagger\phi)^3$ coupling are not renormalized in the $\cN=2$ theory. The diagrams contributing to this correlator again factorize into a product of matter loops, that are in general of a lower order in perturbation theory (the only difference is that the diagrams may now include both $\bar{\psi}\psi\bar{\phi}\phi$ and $(\phi^\dagger\phi)^3$ vertices). The proof then follows in a similar way.

We are now ready to show that $\beta_{\lambda_6} = O(1/N)$.
In our non-supersymmetric model, at large $N$ the correlator $\left< (\cO_1)^3 \right>$ receives contributions from two types of diagrams, with either zero or one $(\phi^\dagger\phi)^3$ vertices:
\begin{center}
\begin{tabular}{cc}
\includegraphics[height=4.5cm]{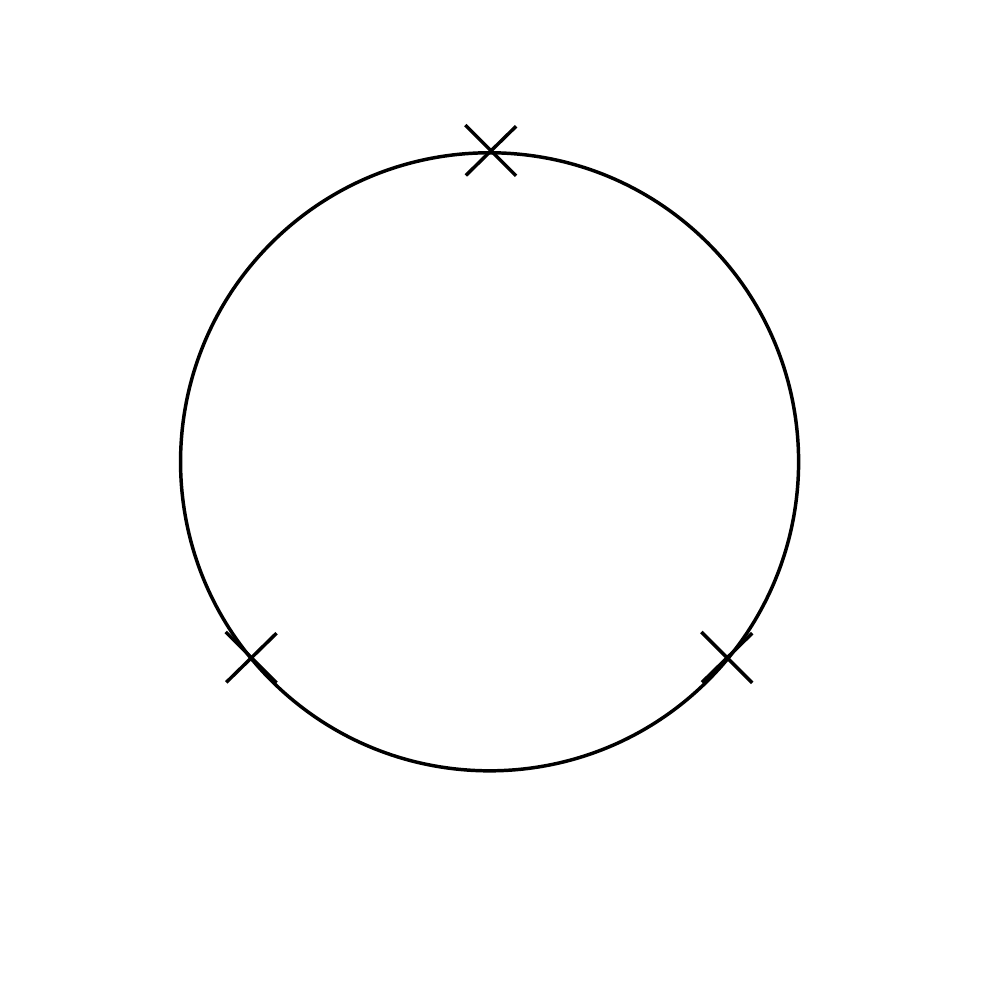} &
\vspace{-2mm}
\includegraphics[height=4.5cm]{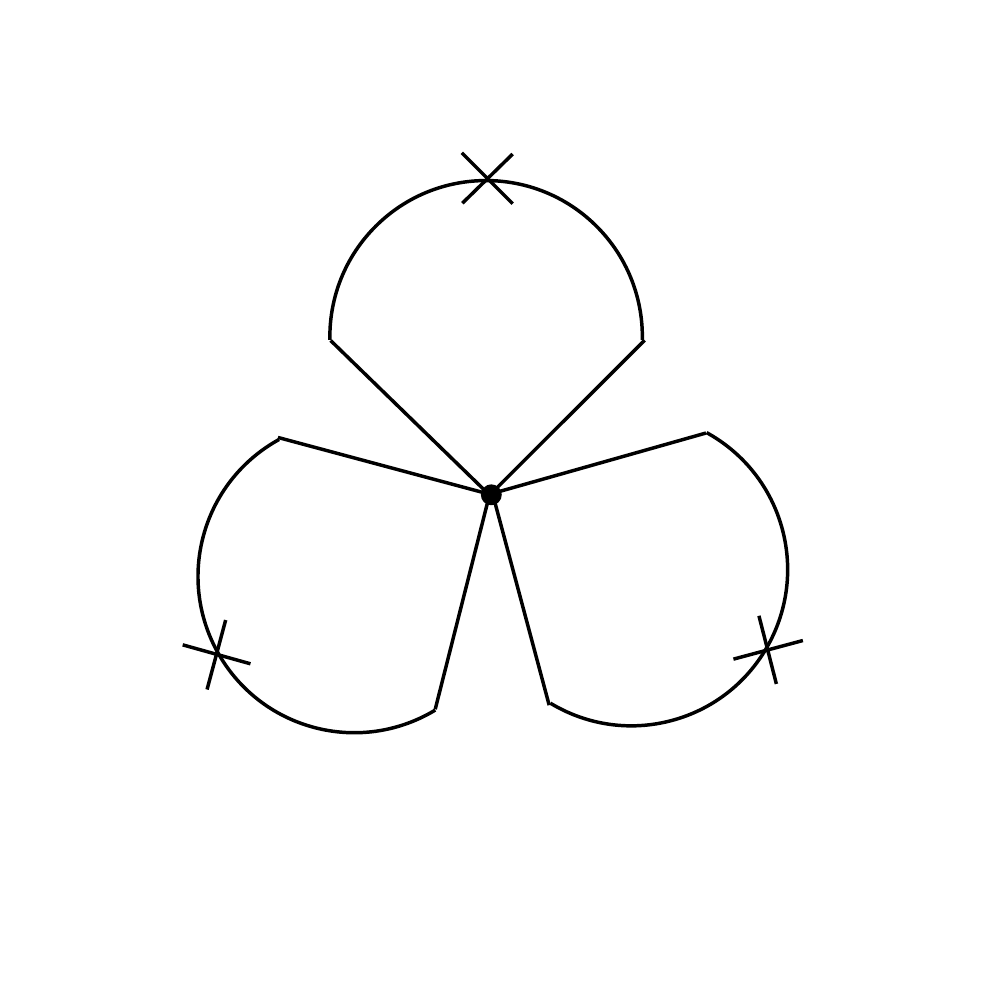}
\vspace{-2mm}
\\
(G1) & (G2)
\end{tabular}
\end{center}
The $\lambda_6$ coupling contributes to $\left< (\cO_1)^3 \right>$ at large $N$ through diagrams of the form (G2), some of which are non-zero (such as the leading order diagram which is explicitly drawn). Thus, if the beta function is non-zero we must have divergences in $\left< (\cO_1)^3 \right>$. However, we have shown above that at every order in $\lambda$ the sum of diagrams (G1) is finite, and also the sum of diagrams (G2) is finite. Thus, the beta function must vanish at large $N$.

\subsection{Spontaneous Breakdown of Conformal Symmetry}
\label{sponbreak}

In order to verify that our theories are conformally invariant, we should also make sure that they do not spontaneously break conformal invariance, by a vacuum expectation value for $\phi^2$. For the theory with $\lambda=0$ and $\lambda_6 \neq 0$, this was analyzed in detail in \cite{Bardeen:1983rv}, and it was found that for $\lambda_6 < (4\pi)^2$ such a breaking does not arise. In fact, the effective potential for $\sigma = \phi^2/N$ can be computed exactly for infinite $N$, and it takes the form \cite{Amit:1984ri}
\begin{equation}
V(\sigma) = \frac{N}{6} \left[ (4\pi)^2 - \lambda_6 \right] |\sigma|^3.
\end{equation}
Thus, for small $\lambda_6$ the only minimum of the effective potential is at the
conformal point $\phi^2=0$. We expect that turning on a small coupling $\lambda$, as we
analyzed above, will lead to small changes in the coefficient of $|\sigma|^3$ in this effective potential (which can be
computed in perturbation theory in $\lambda$), but at least for small $\lambda$ and small $\lambda_6$ it seems clear that there will still be a minimum of the effective potential at
$\sigma=0$. Thus, at least for weak couplings and large $N$, the conformal symmetry is not spontaneously broken in the two-parameter family of conformal field
theories that we discussed above. For $\lambda=0$ a spontaneous breaking of the conformal symmetry can occur when $\lambda_6 = (4\pi)^2$ exactly, and it would be interesting to investigate how this statement is modified at finite $\lambda$ (see \cite{Dias:2003pw,Dias:2010it} for a study of the effective potential in the Abelian Chern-Simons-Matter theory, and \cite{Rabinovici:2011jj} for a similar study of the $O(N)$ vector model with a Chern-Simons term for a $U(1)$ subgroup of $O(N)$). For small values of $N$, spontaneous breaking of the conformal symmetry might happen (as in \cite{Ferrari:2010ex}), and it would be interesting to check if it happens in our theories.

\section{Higher-Spin Currents}
\label{HScurrents}
The main goal of this section is to find the spectrum of
primary operators of the large $N$ interacting fixed points discovered in the previous section. Let us begin by considering the free theory, taking $\lambda=\lambda_6=0$. For each positive, even spin $s$ it has a unique $O(N)$-singlet primary operator $J_s$ that saturates the unitarity bound $\Delta \ge s + d - 2$. (In the theory with a complex scalar in the fundamental representation of $U(N)$ there is such a primary for each positive spin, not just the even ones.) These are symmetric, traceless tensors that can be written schematically as\footnote{We use a normalization in which the 2-point functions remain finite in the large $N$ limit.}
\begin{align}
  J_{\mu_1 \dots \mu_s} &= \frac{1}{\sqrt{N}}
  \phi^i \partial_{\mu_1} \cdots \partial_{\mu_s} \phi^i + \cdots \ed
\end{align}
For example, the first two such operators are
\begin{align}
  J_{\mu\nu} &= \frac{1}{\sqrt{N}} \left\{
  -\frac{1}{3} \phi^i \partial_\mu \partial_\nu \phi^i + \partial_\mu \phi^i \partial_\nu \phi^i - \frac{1}{3} \delta_{\mu\nu} \partial_\rho \phi^i \partial_\rho \phi^i + \frac{1}{9} \delta_{\mu\nu} \phi^i \square \phi^i \right\}
  \ec \label{eq:J2free}\\
  J_{\mu\nu\rho\sigma} &= \frac{1}{\sqrt{N}} \left\{
  \frac{3}{2} \phi^i \partial_\mu \partial_\nu \partial_\rho \partial_\sigma
  \phi^i
  -
  42 \partial_{(\mu} \phi^i
  \partial_\nu \partial_\rho \partial_{\sigma)} \phi^i
  +
  \frac{105}{2} \partial_{(\mu} \partial_\nu \phi^i
  \partial_\rho \partial_{\sigma)} \phi^i
  \right. \nonumber\\ &\quad
  +
  18 \delta_{(\mu\nu} \partial_\rho \partial_{\sigma)}
  \partial_\chi \phi^i \partial_\chi \phi^i
  -
  30 \delta_{(\mu\nu|} \partial_\chi \partial_{|\rho|} \phi^i
  \partial_{\chi} \partial_{|\sigma)} \phi^i
  +
  3 \delta_{(\mu\nu} \delta_{\rho\sigma)}
  \partial_\chi \partial_\xi \phi^i
  \partial_\chi \partial_\xi \phi^i
  \nonumber\\ &\quad
  -
  \frac{9}{7} \delta_{(\mu\nu} \phi^i \partial_\rho \partial_{\sigma)}
  \square \phi^i
  +
  18 \delta_{(\mu\nu} \partial_\rho \phi^i \partial_{\sigma)}
  \square \phi^i
  -
  15 \delta_{(\mu\nu} \partial_\rho \partial_{\sigma)} \phi^i
  \square \phi^i
  \nonumber\\ &\quad \left.
  +
  \frac{9}{70} \delta_{(\mu\nu} \delta_{\rho\sigma)}
  \phi^i \square \square \phi^i
  +
  \frac{3}{2} \delta_{(\mu\nu} \delta_{\rho\sigma)}
  \square \phi^i \square \phi^i
  -
  \frac{18}{5} \delta_{(\mu\nu} \delta_{\rho\sigma)}
  \partial_\chi \phi^i \partial_\chi \square \phi^i \right\} \ec
  \label{eq:J4free}
\end{align}
where parentheses around indices denote an averaging over all permutations of the indices. When discussing the large $N$ limit we shall call such scalar bilinears ``single-trace'' operators. Since they saturate the unitarity bound, these primaries are also conserved currents, $\partial_\mu J^{\mu}_{\;\;\mu_1\dots\mu_{s-1}} = 0$, and therefore the free theory has an infinite number of conserved currents. In addition there is a scalar singlet operator $J_0 = \frac{1}{\sqrt{N}} \phi^i \phi^i$, also a primary, with dimension $\Delta=1$. In the large $N$ limit, all operators in the theory are products of these basic ``single-trace'' operators, or descendants of such products. Note that adding the Chern-Simons sector does not add any additional non-trivial local operators.

Let us now turn on the Chern-Simons coupling $\lambda$. As we showed in section \ref{allorders}, the theory is still conformally-invariant at infinite $N$. The currents of the free theory, as written above, are not gauge-invariant, but they can be made gauge-invariant by promoting derivatives to covariant derivatives and projecting onto the symmetric traceless part. The promoted currents, which will also be denoted $J_s$, are the ``single-trace'' primary operators of the new theory. At finite $N$ they are generally not conserved, and they also mix with ``multi-trace operators''; however, as we shall now see (following a similar analysis in \cite{Girardello:2002pp}) they are still conserved at $N=\infty$.

In the free theory, the primary operator $J_s$ heads a short representation of the conformal group that we label $(\Delta=s+1,s)$, where $\Delta$ is the conformal dimension and $s$ the spin. The shortening condition is the conservation equation $\partial_\mu J^\mu_{\,\,\,\,\,\mu_1\dots\mu_{s-1}} = 0$. For $J_s$ to become non-conserved, there must appear on the right-hand side of this equation a non-zero operator in the representation $(s+2,s-1)$. In other words, $J_s$ must combine with another operator in this representation to form a long representation \cite{Girardello:2002pp},
\begin{align} \label{confreps}
  {\mathrm{lim}}_{\epsilon \to 0}(s+1+\epsilon,s)_{\mathrm{long}} &\cong (s+1,s)_{\mathrm{short}} \oplus (s+2,s-1) \ed
\end{align}
By acting with special conformal transformations on $\partial_\mu J^\mu_{\,\,\,\,\,\mu_1\dots\mu_{s-1}}$ one can show that in the limit in which $J_s$ is conserved, the $(s+2,s-1)$ operator in \eqref{confreps} must be a primary of the conformal algebra \cite{Heidenreich:1980xi} (the coefficient of this operator in the equation for $d*J_s$ vanishes in this
limit, but the special conformal generator acting on $d*J_s$ vanishes even faster).
Now, a connected correlator of the form $\partial_\mu \left< J^\mu_{\,\,\,\,\mu_1\dots\mu_{s-1}} \cO \right>$ can have a leading, $O(N^0)$ contribution in the large $N$ limit only when $\cO$ is a ``single-trace'' operator. Therefore, at $N=\infty$, $J_s$ can only combine with other ``single-trace'' operators. Since there are no such primary operators with $(s+2,s-1)$, $J_s$ must remain conserved even when the Chern-Simons interaction is turned on. Because the representations for conserved currents are short, this also implies that the currents do not acquire an anomalous dimension at this order.

Next we consider the $O(1/\sqrt{N})$ corrections. At this order the currents with $s > 2$ can become non-conserved, but only by combining with a ``double-trace'' operator \cite{Girardello:2002pp} of the schematic form
\begin{align}
  \partial \cdot J_s &\sim
  \frac{f(\lambda)}{\sqrt{N}} \, \epsilon \, \partial^2 J_{s-2} \, J_0 +
  (\mathrm{other\ double\!-\!trace\ operators}) \ec
  \label{noncons}
\end{align}
where $\epsilon$ is the Levi-Civita tensor, and the indices are implicit and can be contracted in various ways. Parity implies that the function $f(\lambda)$
must be odd. Such an equation
implies that $J_s$ has an anomalous dimension of order $1/N$, times some function of $\lambda$.

From \eqref{noncons} it is easy to obtain a non-renormalization theorem for the anomalous dimension of $J_0$ at large $N$ (which we derived by different methods in the previous section).\footnote{We thank S.
Minwalla for discussions on this issue.} By making a scale transformation of \eqref{noncons} and using the fact that $\Delta_s~=~s+1$, we see that the scaling dimension of $J_0$ must be $\Delta_0=1+O(1/\sqrt{N})$, namely it does not get corrections at $N=\infty$, for any value of $\lambda$.
The implicit assumption in this argument is that the coefficient $f(\lambda)$ on the right-hand side of \eqref{noncons} does not vanish. This is indeed what we find for the divergence of (for example) $J_4$ at leading order in $\lambda$ by using the equations of motion,
\begin{align}
  \partial^\sigma J_{\mu\nu\rho\sigma} =
  -\frac{i}{2} \frac{\lambda}{\sqrt{N}} &\left\{
  \frac{540}{7} \,\epsilon_{\alpha\beta(\mu}
  J_{\nu|\alpha}
  \partial_{|\rho)} \partial_\beta J_0
  +
  \frac{396}{7} \,\epsilon_{\alpha\beta(\mu|}
  \partial_\alpha J_{\beta|\nu} \cdot
  \partial_{\rho)} J_0
  -
  \frac{468}{7} \,\epsilon_{\alpha\beta(\mu}
  \partial_\nu J_{\rho)\alpha} \cdot
  \partial_\beta J_0
  \right. \notag\\ &\quad
  -
  \frac{108}{7} \,\epsilon_{\alpha\beta(\mu|}
  \partial_\alpha \partial_{|\nu} J_{\rho)\beta} \cdot
  J_0
  -
  \frac{108}{7} \,\delta_{(\mu\nu} \epsilon_{\rho)\alpha\beta}
  J_{\alpha\chi}
  \partial_\beta \partial_\chi J_0
  \notag\\ &\quad
  -
  \frac{1989}{224} \,\delta_{(\mu\nu} \epsilon_{\rho)\alpha\beta}
  \partial_\alpha J_{\beta\chi} \cdot
  \partial_\chi J_0
  +
  36 \,\epsilon_{\alpha\beta(\mu|}
  \partial_\alpha J_{|\nu\rho)} \cdot
  \partial_\beta J_0
  \notag\\ &\quad \left.
  +
  \frac{3141}{224} \,\delta_{(\mu\nu|} \epsilon_{\alpha\beta\gamma}
  \partial_\alpha J_{|\rho)\beta} \cdot
  \partial_\gamma J_0
  +
  O(\lambda^2,\lambda_6)
  \right\} \ed
  \label{divJ4}
\end{align}
One can verify that the right-hand side of \eqref{divJ4} is traceless at this order, as expected: the traced right-hand side is proportional to
\begin{align}
  \epsilon_{\mu\alpha\beta} \partial_{\alpha} \phi^{i} \partial_{\nu} \partial_{\beta} \phi^{i} \partial_{\nu} \phi^{j} \phi^{j}
   - \epsilon_{\nu\alpha\beta} \partial_{\alpha} \phi^{i} \partial_{\mu} \partial_{\beta} \phi^{i} \partial_{\nu} \phi^{j} \phi^{j}
   + \epsilon_{\mu\alpha\beta} \partial_{\nu} \phi^{i} \partial_{\alpha} \partial_{\nu} \phi^{i} \partial_{\beta} \phi^{j} \phi^{j}
   \ec
\end{align}
and this expression can be shown to vanish by choosing a specific value for
$\mu$ and using the equations of motion.

As another check of \eqref{divJ4} one can act on both sides with $K^\rho$,
the generator of special conformal transformations (see Appendix
\ref{conf} for our conventions). On the left-hand side we have (when the operator is
at $x=0$)
\begin{align}
  [K^\rho, [P^\sigma, J_{\mu\nu\rho\sigma}]] &=
  [[K^\rho, P^\sigma], J_{\mu\nu\rho\sigma}]
  + [P^\sigma, [K^\rho, J_{\mu\nu\rho\sigma}]]
  = 2i [\delta^{\rho\sigma} D + M^{\rho\sigma},
    J_{\mu\nu\rho\sigma}]
  = 0 \ec
\end{align}
where we used the fact that $J_4$ is a primary operator, and that it is symmetric and traceless. The commutator of $K^{\rho}$ with the right-hand side of \eqref{divJ4} should therefore also vanish, and this can be verified directly. The calculation is straightforward, and does not require substituting the explicit expressions for $J_2$ and $J_0$. We have also explicitly verified in appendix~\ref{adim} that $J_0$ indeed has vanishing anomalous dimension at leading order in $1/N$, to two-loop order.

Let us summarize this section. We considered the spectrum of primaries in the 2-parameter family of conformal theories at infinite $N$, found in section~\ref{confSym}. We showed that the spectrum of single-trace, gauge-invariant primaries in these theories is the same as that of the free theory; namely, it consists of conserved higher-spin currents of all even positive spins in the $O(N)$ model (and all positive spins in the $U(N)$ model), plus a scalar operator of conformal dimension 1. For finite $N$, all these operators (except for the conserved energy-momentum tensor $J_2$, and (for the $U(N)$ model) the conserved $U(1)$ current $J_1$) obtain anomalous dimensions.

\section{Correlation Functions}
\label{conjecture}

We have seen above that for infinite $N$ the scaling dimensions in our family of fixed points are independent of $\lambda$, and the deformation of the spectrum is trivial at large $N$. One could then worry that perhaps all correlation functions are independent of $\lambda$. In this section we compute a specific correlation function of currents, $\left<J_2 J_1 J_1\right>$, and show that it does depend on $\lambda$ (already at leading
order in $\lambda$).

One motivation for this computation is to obtain clues towards finding a holographic dual for the theories discussed above. The free theory with $\lambda=\lambda_6=0$ is conjectured \cite{Klebanov:2002ja} to be dual to Vasiliev's higher-spin gauge theory on $AdS_4$, and our theories should be (in the classical limit) continuous deformations of this. The existence of a deformation of Vasiliev's theory, which is dual to the $(\phi^2)^3$ deformation of the free vector model, was first mentioned in \cite{Elitzur:2005kz}. For that deformation the holographic picture is clear, since this is a ``multi-trace'' deformation that is manifested in the holographic dual as a change in boundary conditions of the scalar field dual to $\phi^2$ \cite{Witten:2001ua,Berkooz:2002ug,Elitzur:2005kz}.

On the other hand, the Chern-Simons deformation by $\lambda$ should be realized on the gravity side as a continuous, parity-breaking deformation of Vasiliev's theory. One natural conjecture could be that it is dual to one of the known parity-breaking deformations of Vasiliev's theory, which were parameterized in \cite{Sezgin:2002rt} by some odd function ${\cal V}(X)$. However, as mentioned in \cite{Giombi:2011kc}, such a deformation seems not to lead to a non-vanishing $\left<J_2 J_1 J_1\right>$ at leading order in $\lambda$ as we find below. If so, there should be some new, unknown deformation of Vasiliev's theory that is dual to turning on $\lambda$, and it would be very interesting to discover it.

\subsection{Computation of $\left<J_2 J_1 J_1\right>$}
\label{J2J1J1}

Corrections to correlation functions at order $\lambda$ necessarily break parity.
For simplicity, we study here the $U(N)$ case, which has a conserved current $J_1$,
since the correlator $\left<J_2 J_1 J_1\right>$ is the simplest correlator of conserved currents that can exhibit a parity-breaking structure \cite{Giombi:2011rz}.
The conserved currents $J=J_1$, $T=J_2$ of the theory of $N$ complex scalars with $U(N)$ Chern-Simons interactions are given by
\begin{align}
  J_\mu &= \frac{1}{\sqrt{N}}\left\{i (D_\mu \phi)^\dagger \phi - i \phi^\dagger D_\mu \phi\right\}
  \ec
  \label{J1cpx}
  \\
  T_{\mu\nu} &= \frac{1}{\sqrt{N}} \left\{
  -\frac{1}{6} ( \phi^\dagger D_\mu D_\nu \phi
  + \phi^\dagger D_\nu D_\mu \phi
  + D_\mu D_\nu \phi^\dagger \cdot \phi
  + D_\nu D_\mu \phi^\dagger \cdot \phi )
  + D_\mu \phi^\dagger D_\nu \phi
  + D_\nu \phi^\dagger D_\mu \phi
  \notag \right.\\ &\quad \left.
  \qquad \quad
  - \frac{2}{3} \delta_{\mu\nu} D_\rho \phi^\dagger D_\rho \phi
  + \frac{1}{9} \delta_{\mu\nu} \phi^\dagger D^2 \phi
  + \frac{1}{9} \delta_{\mu\nu} D^2 \phi^\dagger \cdot \phi \right\}
  \ec
  \label{emtensorcpx}
\end{align}
where $D_\mu = \partial_\mu + g A_\mu^a T^a$.

\newcommand{\pol}{\varepsilon}

With these definitions the 2-point functions of $J$ and $T$ in the free theory are fixed to be (denoting e.g. $J_{\pol}(x) = \pol^{\mu} J_{\mu}(x)$)
\begin{align}
\langle J_{\pol_1}(x_1) J_{\pol_2}(x_2) \rangle &= \frac{1}{8\pi^2|x_{12}|^2}\pol_1^{\mu}\pol_2^{\nu}\left(\frac{\delta^{\mu\nu}}{|x_{12}|^2}-2\frac{x_{12}^{\mu}x_{12}^{\nu}}{|x_{12}|^4}\right) \ec\\
\langle T_{\pol_1\pol_1}(x_1) T_{\pol_2\pol_2}(x_2) \rangle &= \frac{1}{3\pi^2|x_{12}|^2}\left[\pol_1^{\mu}\pol_2^{\nu}\left(\frac{\delta^{\mu\nu}}{|x_{12}|^2}-2\frac{x_{12}^{\mu}x_{12}^{\nu}}{|x_{12}|^4}\right)\right] ^2 \ed
\end{align}

We now compute the correlator $\left< T_{\pol_1 \pol_2}(x_1) J_{\pol_3}(x_2) J_{\pol_4}(x_3) \right>$ in $x$-space at order $\lambda$. It has a unique parity-violating tensor structure, and to compute its coefficient it will prove useful (as in \cite{Giombi:2011kc}) to choose all polarizations equal and null, $\pol_i = \pol$, $\pol^2 = 0$, and to take the limit $x_2 \to x_1$. With these choices, the parity-violating tensor structure has the form \cite{Giombi:2011rz}
\begin{align}
  \frac{1}{|x_{12}||x_{23}||x_{13}|} \left(
  Q_1^2 S_1 + 2 P_2^2 S_3 + 2 P_3^2 S_2 \right) \to
  -\frac{4 \epsilon_{\mu\nu\rho} x_{13}^\mu x_{12}^\nu \pol^\rho
  (\pol\cdot x_{12})^2 (\pol\cdot x_{13}) }
  { |x_{12}|^6 |x_{13}|^6 } \ed
  \label{J2J1J1struct}
\end{align}
In the limit $x_2 \to x_1$ it diverges as $|x_{12}|^{-3}$, and we shall use this fact to discard subleading terms in $|x_{12}|$.

There are 3 diagrams, up to permutations of the current insertions, contributing to the correlator at order $\lambda$:

\begin{center}
\begin{tabular}{ccc}
  \hspace{-1.5cm}
\includegraphics[width=5.2cm]{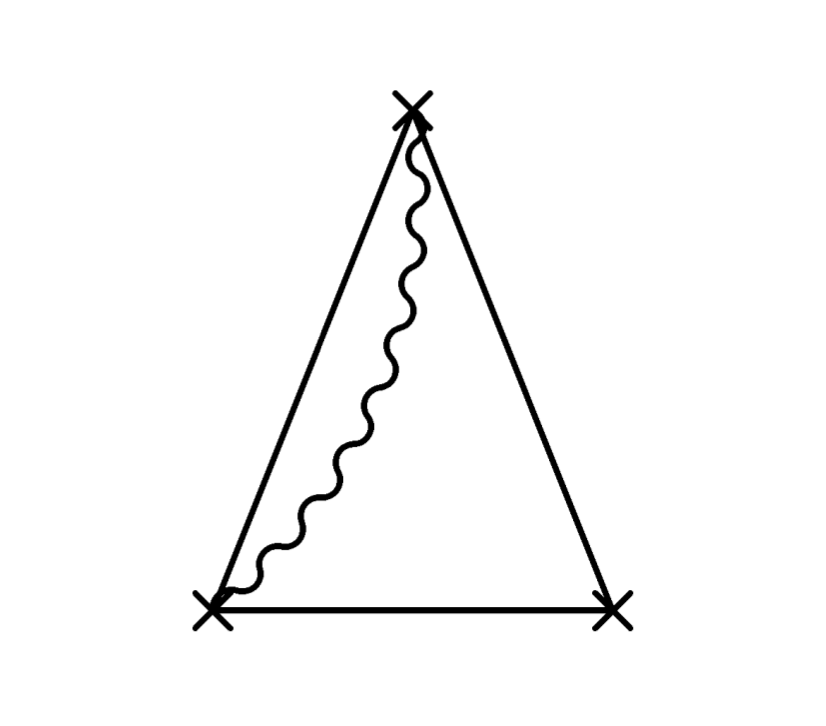} &
\includegraphics[width=5.2cm]{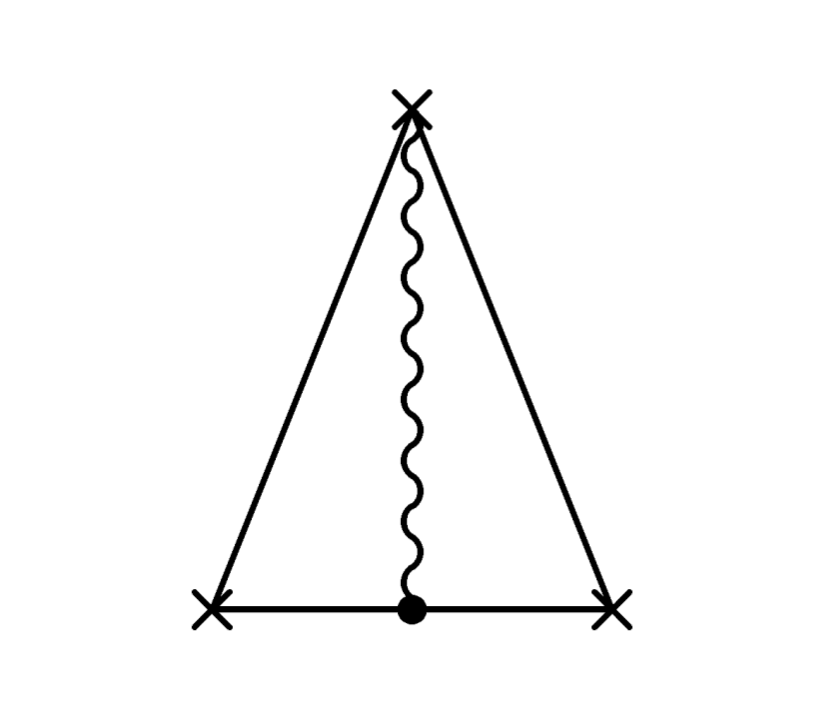} &
\includegraphics[width=5.2cm]{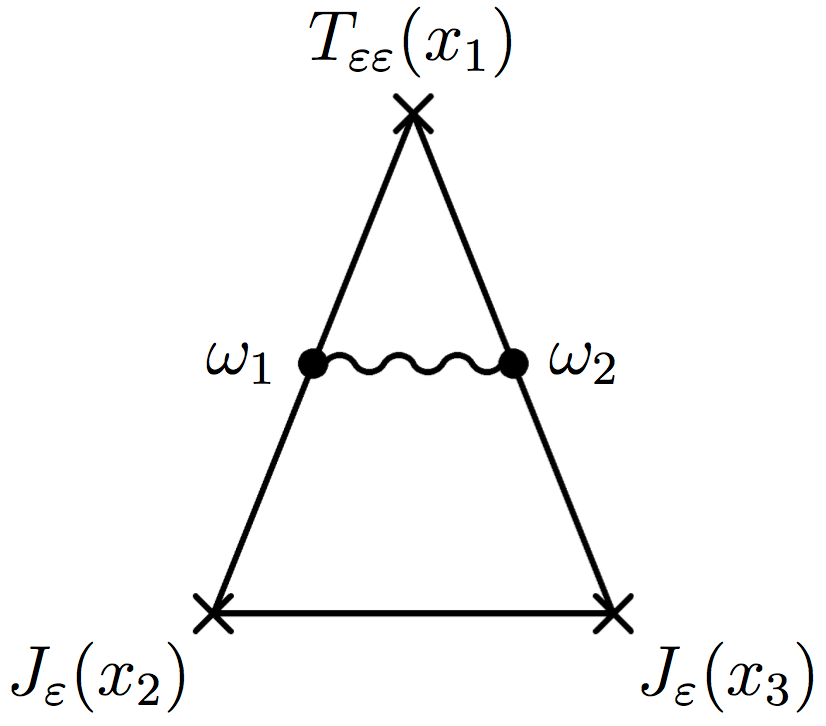}
\\
  \hspace{-1.5cm}
(D1) & (D2) & (D3)
\end{tabular}
\end{center}

Diagrams of the type (D1) vanish because all polarizations are equal: from the gluon propagator we have $\epsilon_{\mu\nu\rho} \varepsilon^\mu \varepsilon^\nu = 0$. The computation of (D2), (D3) is not completely straightforward; it involves the repeated application of several techniques, as we will demonstrate by computing one of the (D3) diagrams in detail. The results for the other diagrams are listed in Appendix \ref{diags}.

In this theory the scalar and gluon propagators are given by
\begin{align}
I_{xy} &= \frac{1}{4 \pi}\frac{1}{|x-y|} \ecq
I_{\mu\nu;xy} = -\frac{i}{4\pi}\frac{\epsilon_{\mu\nu\rho}(x-y)^{\rho}}{|x-y|^3}\ed
\end{align}

To regularize the diagrams we change the loop variables to be $d$ dimensional:  $d^3\om \to d^d\om$. While this is not a gauge-invariant regulator, we found that each of the diagrams is finite and thus independent of $d$. We have also checked that using a different regulator gives the same results.

The diagram (D3), with the gluon line stretched between the two scalar propagators connected to the energy-momentum tensor, is given by
\begin{align}
-2 \frac{\lambda}{\sqrt{N}}
\int \! d^d\om_1 \, d^d\om_2 \, I_{\alpha\beta;\om_1\om_2}
\left[ I_{\om_2 x_1} T^0_{\e\e}(x_1) I_{x_1\om_1}
\olra{\p}_{\!\!\om_1,\alpha} I_{\om_1 x_2}
(\pol\cdot\olra{\p}_{\!\!x_2}) I_{x_2 x_3}
(\pol\cdot\olra{\p}_{\!\!x_3}) I_{x_3\om_2}
\olra{\p}_{\!\!\om_2,\beta} \right]
\ed \label{eq:D3Perm1}
\end{align}
Here $\olra{\p}\equiv \overrightarrow{\p}-\overleftarrow{\p}$, and
\begin{align}
  \phi^{i\dag}(x_1)T^0_{\pol\pol}(x_1)\phi^i(x_1)\equiv
  \phi^{i\dag}(x_1)\left[-\frac{1}{3}(\pol\cdot\ora{\p_{x_1}})^2-\frac{1}{3}(\pol\cdot\ola{\p_{x_1}})^2+
  2(\pol\cdot\ola{\p_{x_1}})(\pol\cdot\ora{\p_{x_1}})\right]\phi^{i}(x_1)
\end{align}
is the energy-momentum tensor at leading order. It is understood that the right-most derivative $\overrightarrow{\p}$ in \eqref{eq:D3Perm1} acts on the left-most propagator inside the brackets.

Let us try and take as many derivatives as possible out of the integral. We are limited by the fact that there are two propagators involving $x_1$, and the combination $T^0_{\pol\pol}$ is not a total derivative acting on them. To proceed let us first split the point $x_1$ into two points $x_1,x_1'$, each connected to a different scalar line; eventually we will take $x_1' \to x_1$. This procedure does not spoil gauge invariance, since at this order in $\lambda$ it is equivalent to stretching a Wilson line between the separated points. The result can be written as
\begin{align}
  \frac{2i \lambda}{(4\pi)^6\sqrt{N}}\frac{1}{|x_{23}|}
  \left. (\pol\cdot\olra{\p_{x_2}})
  (\pol\cdot\olra{\p_{x_3}})
  \left(\p_{x_2,\alpha}-\p_{x_1,\alpha}\right) \cal{I}_\alpha \, \right|_{x_1'\rightarrow x_1} \ec
  \label{D3simp}
\end{align}
where $x_{ij} \equiv x_i - x_j$, and
\begin{align}
  \cal{I}_\alpha &= 2 \int \! d^d\om_1 \, d^d\om_2 \,
  \frac{\e_{\alpha\beta\gamma}\om_{12}^{\gamma}(\om_2-x_3)^{\beta}}{|\om_{12}|^3|\om_1-x_1||\om_2-x_1'||\om_1-x_2||\om_2-x_3|^3} \times \notag\\ &\quad \qquad
  \times \left[ 2\frac{(\pol\cdot(\om_1-x_1))(\pol\cdot(\om_2-x_1'))}{|\om_1-x_1|^2|\om_2-x_1'|^2} -\frac{(\pol\cdot(\om_1-x_1))^2}{|\om_1-x_1|^4} - \frac{(\pol\cdot(\om_2-x_1'))^2}{|\om_2-x_1'|^4} \right]
\label{eq:A10} \ed
\end{align}
To arrive at this form, we used the relation $(\om_2-x_3)^{\beta} / |\om_2-x_3|^3 =-\p_{\om_2}^{\beta} |\om_2-x_3|^{-1}$, and integrated by parts with respect to $\om_2$. Note that we have chosen to take out a single $x_1$ derivative, while acting with the rest explicitly.

Next, note that $\p_{x_1',\alpha} \cal{I}_\alpha = 0$, as can be seen by rewriting $\p_{1'}^{\alpha}$ as $\p_{\om_2}^{\alpha}$ and integrating by parts. This means that we can take $x_1'\rightarrow x_1$ before acting with the outer derivatives in \eqref{D3simp}, since the $x_1$ derivative there acts in the $\alpha$ direction. In addition, let us shift $\om_{1,2} \to \om_{1,2} + x_1$. With these changes, the integral simplifies to
\begin{align}
  \cal{I}_\alpha &= 2\int \! d^d\om_1 \, d^d\om_2 \, \frac{\e_{\alpha\beta\gamma} \om_{12}^{\gamma}(\om_2+x_{13})^{\beta}}{|\om_{12}|^3|\om_1||\om_2||\om_1+x_{12}||\om_2+x_{13}|^3}
  \left[
  2\frac{(\pol\cdot\om_1)(\pol\cdot\om_2)}{|\om_1|^2|\om_2|^2}
  - \frac{(\pol\cdot\om_2)^2}{|\om_2|^4}
  - \frac{(\pol\cdot\om_1)^2}{|\om_1|^4}
  \right]\ed
  \label{eq:A1}
\end{align}

Let us now consider the limit $x_2 \to x_1$, in which the integral \eqref{eq:A1} diverges as $|x_{12}|^{-1}$. In this limit, the full diagram \eqref{D3simp} diverges as $|x_{12}|^{-3}$, and therefore contributes to the parity-violating tensor structure \eqref{J2J1J1struct}. We first compute the last term in the brackets in \eqref{eq:A1}. Using the fact that $\om_{12}^\gamma/|\om_{12}|^3 = - \partial_{\om_2,\gamma} |\om_{12}|^{-1}$ and integrating by parts we rewrite this term as
\begin{align}
  \cal{I}_\alpha^{\text{last~term}} &= -2\int \! d^d\om_1 \, d^3\om_2 \, \frac{ (\pol\cdot\om_1)^2\e_{\alpha\beta\gamma}\om_2^{\gamma}x_{13}^{\beta} }{ |\om_{12}||\om_1|^5|\om_2|^3|\om_1+x_{12}||\om_2+x_{13}|^3} \nonumber\\
&= -\frac{1}{\pi}\int \! d^d\om_1 \, d^5\om_2 \, \frac{ (\pol\cdot\om_1)^2\e_{\alpha\beta\gamma}\om_1^{\gamma}x_{13}^{\beta} }{ |\om_{12}|^3|\om_1|^5|\om_2|^3|\om_1+x_{12}||\om_2+x_{13}|^3} \ed
\end{align}
The second equality can be verified by introducing Feynman parameters and performing the dimensional integration on both sides. The integral over $\om_2$ can now be carried out \cite{Boos:1987bg}, and we find
\begin{align}
  \cal{I}_\alpha^{\text{last~term}} = -\frac{8\pi}{|x_{13}|}\int \! d^3\om_1 \, \frac{(\pol\cdot\om_1)^2\e_{\alpha\beta\gamma}\om_1^{\gamma}x_{13}^{\beta}}{|\om_1|^6|\om_1+x_{12}||\om_1+x_{13}|}\frac{1}{|\om_1|+|x_{13}|+|\om_1+x_{13}|} \ed
\end{align}
As mentioned above, this integral diverges as $|x_{12}|^{-1}$ in the limit $x_2 \to x_1$, and the divergence comes from the region $|\om_1| \ll 1$. As we approach the limit, most of the contribution to the integral will therefore come from this region. We can therefore expand around $\om_1 = 0$, keeping only the leading term; the remaining terms will give sub-leading corrections in $|x_{12}|$. We thus arrive at a straightforward integral,
\begin{align}
  \cal{I}_\alpha^{\text{last~term}} = -4\pi\frac{x_{13}^{\beta}}{|x_{13}|^3}\int \! d^3\om_1 \, \frac{(\pol\cdot\om_1)^2\e_{\alpha\beta\gamma}\om_1^{\gamma}}{|\om_1|^6|\om_1+x_{12}|}
  + O((x_{12})^0) \ed
  \label{Alast}
\end{align}

The other two terms in \eqref{eq:A1} have an $|x_{12}|^{-1}$ divergence in the limit $x_2 \to x_1$ coming from the region $|\om_1|,|\om_2| \ll 1$, and we can similarly take the leading order in the expansion around $\om_{1,2} = 0$. The resulting integral is again straightforward to evaluate,
\begin{align}
  \cal{I}_\alpha^{\text{terms~1,2}} &= 2\frac{x_{13}^{\beta}}{|x_{13}|^3}
  \int \! d^d\om_1 \, d^d\om_2 \,
  \frac{\e_{\alpha\beta\gamma}\om_{12}^{\gamma}}{|\om_{12}|^3|\om_1||\om_2||\om_1+x_{12}|}
  \left[
  2\frac{(\pol\cdot\om_1)(\pol\cdot\om_2)}{|\om_1|^2|\om_2|^2}
  - \frac{(\pol\cdot\om_2)^2}{|\om_2|^4} \right]
  + O((x_{12})^0) \ed \label{eq:A12}
\end{align}

Combining the results of \eqref{Alast} and \eqref{eq:A12} and plugging into \eqref{D3simp}, the contribution of the specific (D3) diagram that we computed to the parity-violating tensor structure is
\begin{align}\label{eq:D3Perm1Res}
  \frac{i}{24\pi^4}\frac{\lambda}{\sqrt{N}}\frac{(\pol\cdot x_{12})^2(\pol\cdot x_{13})\e_{\alpha\beta\gamma}x_{12}^{\alpha}x_{13}^{\beta}\pol^{\gamma}}{|x_{13}|^6|x_{12}|^6}
  \ed
\end{align}

By applying similar techniques one can compute the other (D2), (D3) diagrams and their permutations; the results are listed in Appendix \ref{diags}. Summing these contributions, we find the following non-zero result at order $\lambda$,
\begin{align}
  \left. \langle T_{\pol\pol}(x_1)J_{\pol}(x_2)J_{\pol}(x_3)\rangle \right|_{x_2 \to x_1} = \frac{i}{24\pi^4}\frac{\lambda}{\sqrt{N}}\frac{(\pol\cdot x_{12})^2(\pol\cdot x_{13})\e_{\alpha\beta\gamma}x_{12}^{\alpha}x_{13}^{\beta}\pol^{\gamma}}{|x_{13}|^6|x_{12}|^6}.
\end{align}
Using the known tensor structure \eqref{J2J1J1struct}, for general coordinates and polarizations this implies
\begin{align}
\langle T_{\pol_1\pol_1}(x_1)J_{\pol_2}(x_2)J_{\pol_3}(x_3)\rangle &= \frac{i}{96\pi^4}\frac{\lambda}{\sqrt{N}}\frac{1}{|x_{12}||x_{23}||x_{13}|} \left(
  Q_1^2 S_1 + 2 P_2^2 S_3 + 2 P_3^2 S_2 \right)
  + o(\lambda^2) \ed
\end{align}

\section{Summary and Future Directions}
\label{summary}

In this paper we studied the three dimensional $O(N)$ ($U(N)$) vector model coupled to a Chern-Simons theory at level $k$, in the limit of large $N,k$ with a fixed ratio $\lambda = 4 \pi N / k$. We found that for infinite $N$ this theory has two exactly marginal deformations, corresponding to $\lambda$ and to a $(\phi^2)^3$ coupling, while for finite large $N$ we showed that there is (at least for small $\lambda$) a single IR-stable fixed point for every $\lambda$. For infinite $N$ we showed
that none of the operators of the theory have anomalous dimensions, so that the infinite tower
of conserved currents of the theory with $\lambda=0$ remains also for finite $\lambda$ (and finite $\lambda_6$). We
showed explicitly that some of the correlation functions of the infinite $N$ theory do depend
on $\lambda$.

The fact that at infinite $N$ we find an infinite tower of conserved currents even in the
interacting theory at finite $\lambda$ is quite surprising, and suggests that this theory may
have some interesting integrable structure. In this paper we only performed explicit computations at low orders in perturbation theory. However, the existence of an infinite number of
conserved currents may be useful towards performing exact computations as a function of $\lambda$
in these theories.
When our scalar fields are replaced by fermion fields, many such exact computations can indeed be performed \cite{Giombi:2011kc}. In this case there is a choice of gauge for which only rainbow diagrams contribute, simplifying the resummation of all planar diagrams. For scalar fields we have not yet been able to find similar simplifications.

Vector models of the type we analyze here exhibit large $N$ phase transitions at temperatures of order $\sqrt{N}$  \cite{Shenker:2011zf,Giombi:2011kc}. It would be interesting to generalize these transitions to our finite $\lambda$ theories.

It would also be interesting to understand the holographic duals of the theories with finite
$\lambda$ that we discussed here, which should be continuous deformations of Vasiliev's
higher-spin theories. Unlike standard marginal deformations, here we are not deforming by the integral of a gauge-invariant local operator, so it is not obvious how to identify this deformation. Perhaps the attempted derivations of the $AdS$/CFT correspondence
for $\lambda=0$ \cite{Das:2003vw,Koch:2010cy,Douglas:2010rc,Jevicki:2011ss} can be generalized to finite $\lambda$, by replacing the scalar
bilinear operators $\phi_i(x) \phi_i(y)$ appearing in these derivations by a gauge-invariant
bilinear (in which the two scalars are connected by an open Wilson line); if so then this
should provide clues towards the construction of this holographic dual.

It would also be
interesting to understand finite $N$ corrections to our theories on the gravity side, though
this may require a quantum completion of Vasiliev's higher-spin theory that is not yet known.
Since on the field theory side our theories are vector models, it seems that they should not
correspond to closed string theories, but to open string theories coupled to a trivial
(topological) closed string background. For instance, since the closed string duals of the
$O(N)_k$ and $U(N)_k$ Chern-Simons theories are known topological string theories \cite{Gopakumar:1998ki,Sinha:2000ap}, one
could imagine that adding fundamental matter fields to these theories (as we have done) should
correspond to adding non-topological D-branes to these topological string theories.

There are many possible generalizations of our computations. The generalization to the
case of $l$ vectors of scalar fields is straightforward, and all the operators we
discuss just become $l\times l$ matrices (the description of this on the gravity side is
straightforward). The anomalous dimensions of all these operators
still vanish in the large $N$ limit, so in particular we have many massless ``gravitons'' in this case, as expected for a theory involving $l$ D-branes. The generalization to fermionic fields instead of scalars will be discussed in \cite{Giombi:2011kc}. One can also consider an ${\cN=1}$ supersymmetric generalization of our theories, whose field content includes both a scalar and a fermionic field, with specific interactions between them. The gravity dual for this case was discussed in \cite{Leigh:2003gk,Sezgin:2003pt}, and it would be interesting to generalize our discussion of the theory with finite $\lambda$ to this case. It would also be interesting to find the gravity dual for the ${\cN=2}$ generalizations of our theories, that we briefly discussed in \S\ref{allorders}.

We hope that further study of these theories will shed more light on the structure of the
$AdS$/CFT correspondence in the case where it gives a weak-weak coupling duality, and hopefully also
in general.

\subsection*{Acknowledgments}
\label{s:acks}

We would like to thank R. Gopakumar and S. Minwalla for interesting discussions that initiated this project, and S. Giombi and S. Minwalla for many useful discussions and for notifying us of the results of \cite{Giombi:2011kc}. We also thank D. Jafferis, Z. Komargodski, E. Rabinovici, M. Smolkin, S. Wadia and S. Yankielowicz for useful
discussions.
This work was supported in part by the Israel--U.S.~Binational Science Foundation, by a research center supported by the Israel Science Foundation (grant number 1468/06), by the German-Israeli Foundation (GIF) for Scientific Research and Development, and by the Minerva foundation with funding from the Federal German Ministry for Education and Research.

\appendix

\section{Conventions}
\label{conventions}

Starting with the action \eqref{eq:action}, let us separate it to the physical coupling part plus counterterms, $\delta Z_x = Z_x - 1$, $\delta\alpha = \frac{1}{2\gamma_R} - \frac{1}{2\alpha}$, so that
\begin{align}
  S &=
  S_\mathrm{CS}^\mathrm{phys.}
  + S_\mathrm{gh}^\mathrm{phys.}
  + S_\mathrm{b}^\mathrm{phys.}
  + S_\mathrm{CS}^\mathrm{c.t.}
  + S_\mathrm{gh}^\mathrm{c.t.}
  + S_\mathrm{b}^\mathrm{c.t.} \, ,\\\nonumber
  \\
  S_\mathrm{CS}^\mathrm{phys.}  &= \int \! d^dx\, \left\{
  - \frac{i}{2}
  \epsilon_{\mu\nu\lambda} A_\mu^a \partial_\nu A_\lambda^a
  - \frac{i}{6} \mu^{\epsilon/2} g \epsilon_{\mu\nu\lambda} f^{abc}
  A_\mu^a A_\nu^b A_\lambda^c
  \right\} \ec\\
  S_\mathrm{gh}^\mathrm{phys.}  &= \int \! d^dx\, \left\{
  - \frac{1}{2 \alpha} (\partial_\mu A_\mu^a)^2
  + \partial_\mu \bar{c}^a \partial^\mu c^a
  + \mu^{\epsilon/2} g f^{abc}
  \partial_\mu \bar{c}^a A_\mu^b c^c
  \right\} \ec\\
  S_\mathrm{b}^\mathrm{phys.}  &= \int \! d^dx\, \left\{
  \frac{1}{2} (\partial_\mu \phi_i)^2
  + \mu^{\epsilon/2} g \partial_\mu \phi_i T_{ij}^a
  A_\mu^a \phi_j
  - \frac{1}{4} \mu^\epsilon g^2 \{T^a,T^b\}_{ij}
  \phi_i \phi_j A_\mu^a A_\mu^b
  \right. \notag\\ &\quad \left. \qquad \qquad
  + \mu^{2\epsilon} \frac{g_6}{3!\cdot 2^3}(\phi_i\phi_i)^3
  \right\} \ec \\\nonumber
  \\
  S_\mathrm{CS}^\mathrm{c.t.} &= \int \! d^dx\, \left\{
  - \frac{i}{2} \delta Z_A
  \epsilon_{\mu\nu\lambda} A_\mu^a \partial_\nu A_\lambda^a
  - \frac{i}{6} \mu^{\epsilon/2} g \delta Z_g \epsilon_{\mu\nu\lambda} f^{abc}
  A_\mu^a A_\nu^b A_\lambda^c
  \right\} \ec\\
  S_\mathrm{gh}^\mathrm{c.t.} &= \int \! d^dx\, \left\{
  - \delta\alpha (\partial_\mu A_\mu^a)^2
  + \delta Z_\mathrm{gh} \partial_\mu \bar{c}^a \partial^\mu c^a
  + \mu^{\epsilon/2} \delta\tilde{Z}_g g f^{abc}
  \partial_\mu \bar{c}^a A_\mu^b c^c
  \right\} \ec\\
  S_\mathrm{b}^\mathrm{c.t.} &= \int \! d^dx\, \left\{
  \frac{1}{2} \delta Z_\phi (\partial_\mu \phi_i)^2
  + \mu^{\epsilon/2} \delta Z_g' g \partial_\mu \phi_i T_{ij}^a
  A_\mu^a \phi_j
  - \frac{1}{4} \mu^\epsilon \delta Z_g'' g^2 \{T^a,T^b\}_{ij}
  \phi_i \phi_j A_\mu^a A_\mu^b \right. \nonumber\\
  &\quad \qquad \qquad
  \left.+ \mu^{2\epsilon} \delta Z_{g_6}\frac{g_6}{3!\cdot 2^3}(\phi_i\phi_i)^3
  \right\}
  \ed
\end{align}
We use Landau gauge, $\alpha\to 0$, in which the gluon propagator is
\begin{align}
\label{eq:GluonProp}
  - \delta_{ab} \epsilon_{\mu\nu\lambda} \frac{p^\lambda}{p^2} \ed
\end{align}
The $O(N)$ generators in the fundamental are taken to be real and anti-symmetric, $(T^a)^{\dag}=(T^a)^T=-T^a$. They satisfy
\begin{gather}
  \trace\left(T^aT^b\right)=\delta^{ab}C_1 \,,\quad
  f^{acd}f^{bcd}=\delta^{ab}C_2 \,,\quad
  T^a_{ij}T^a_{kl}=I_{ij,kl}C_3 \,,\quad \notag\\
  f^{abc} T^b_{ik} T^c_{kj} = \frac{1}{2} C_2 T^a_{ij} \, ,\quad
  f^{abc} = \trace \left( T^a [T^b,T^c] \right)
  \ec
  \label{eq:TrRel}
\end{gather}
where
\begin{align}\label{eq:TrRelOn}
C_1=C_3=1 \ecq
I_{ij,kl}=\frac{1}{2}\left(\delta_{il}\delta_{kj}-\delta_{ik}\delta_{jl}\right) \ecq
C_2=2-N \ed
\end{align}

We will also be interested in the case of a complex scalar field in the fundamental representation of $U(N)$, again coupled to gauge fields with a Chern-Simons interaction. In this case the scalar action is
\begin{align}
S_\mathrm{b} &= \int \! d^dx \,
  \left\{
    Z_\phi |\cal{D}_{\mu} \phi_i|^2
    + \mu^{2\epsilon} Z_{g_6} \frac{g_{6}}{3!}
    (\phi^\dagger \phi)^3
  \right\} \ec
  \label{eq:cSb}
\end{align}
and the generators of $U(N)$ in the fundamental representation satisfy \eqref{eq:TrRel}, with
\begin{align}\label{eq:TrRelUn}
C_1=C_3=1 \ecq
I_{ij,kl}=\delta_{il}\delta_{kj}
\ecq
C_2=-2N \ed
\end{align}
The $SU(N)$ case is identical at large $N$, differing by an extra term in $I_{ij,kl}$. The counterterms for the complex and real theories are related by
\begin{align}
  \delta Z^{SU(N)}_{\phi}&= 4\delta Z^{O(N)}_{\phi} \ecq
  \delta Z^{SU(N)}_{g_6} = 4\delta Z^{O(N)}_{g_6} \ed
  \label{eq:cScalarRen}
\end{align}

\section{2-Loop Diagram Results}
\label{diags}

The following are the diverging parts of the diagrams of the $O(N)$ Chern-Simons-matter theory appearing in sections \ref{2loop} and \ref{J2J1J1}, and in appendix \ref{adim}.

\begin{align}
  (\mathrm{A}1) &= -g^4g_6\deltaperms\left(\frac{3}{2}N^2+\frac{21}{2}N-12\right)\frac{1}{64\pi^2}\frac{1}{\epsilon}\ec \\
  (\mathrm{A}2) &= g^8\deltaperms\left(N^2+N-2\right)\frac{3}{64\pi^2}\frac{1}{\epsilon}\ec \\
  (\mathrm{A}3) &= -g^8\deltaperms\left(N^2-3N+2\right)\frac{3}{64\pi^2}\frac{1}{\epsilon}\ec \\
  (\mathrm{A}4) &= -g^4g_6\deltaperms\left(N-1\right)\frac{9}{32\pi^2}\frac{1}{\epsilon}\ec \\
  (\mathrm{A}5) &= g^8\deltaperms\left(N-1\right)\frac{3}{64\pi^2}\frac{1}{\epsilon}\ec \\
  (\mathrm{A}6) &= 0\ec \\
  (\mathrm{A}7) &= g^8\deltaperms\left(N-1\right)\frac{9}{32\pi^2}\frac{1}{\epsilon}\ec \\
  (\mathrm{A}8) &= g_6^2\deltaperms\left(3N+22\right)\frac{1}{32\pi^2}\frac{1}{\epsilon}\ed
\end{align}
\begin{align}
  (\mathrm{B}1) &= -g^4\delta_{ij}p^2\left(N^2-3N+2\right)\frac{1}{96\pi^2}\frac{1}{\epsilon}\ec \\
  (\mathrm{B}2) &= g^4\delta_{ij}p^2\left(N^2-N\right)\frac{1}{384\pi^2}\frac{1}{\epsilon}\ec \\
  (\mathrm{B}3) &= g^4\delta_{ij}p^2\left(N-1\right)\frac{1}{48\pi^2}\frac{1}{\epsilon}\ec \\
  (\mathrm{B}4) &= g^4\delta_{ij}p^2\left(N-1\right)\frac{1}{96\pi^2}\frac{1}{\epsilon}\ed
\end{align}
\begin{align}
  (\mathrm{C}1) &= g^4\delta_{i_1i_2}\left(\frac{3}{2}N^2+\frac{21}{2}N-12\right)\frac{1}{96\pi^2}\frac{1}{\epsilon} \ec\\
  (\mathrm{C}2) &= g^4\delta_{i_1i_2}\left(N-1\right)\frac{3}{16\pi^2}\frac{1}{\epsilon} \ed
\end{align}

Let us denote the diagrams (D2),(D3) of section \ref{J2J1J1}, including permutations, as

\begin{center}
\begin{tabular}{ccc}
\includegraphics[height=4.5cm]{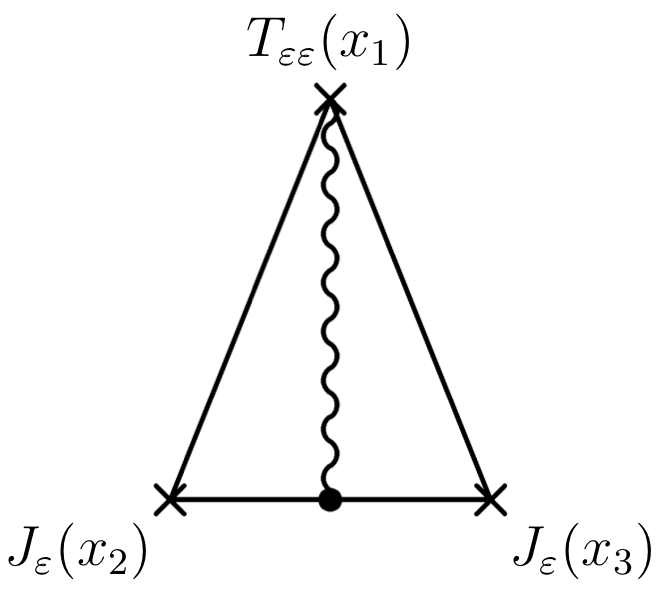} &
\includegraphics[height=4.5cm]{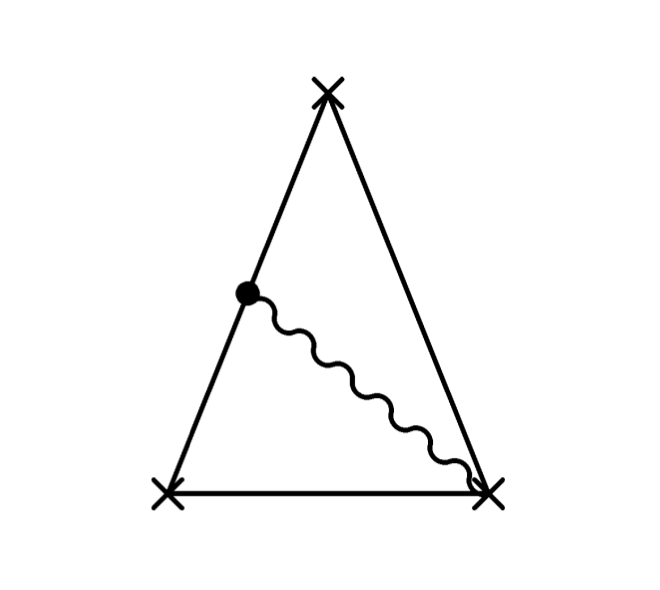} &
\includegraphics[height=4.5cm]{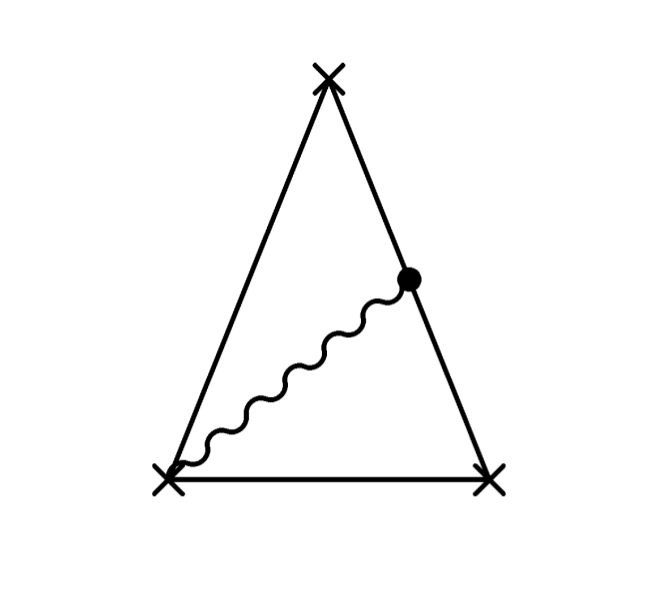}
\vspace{-4mm}
\\
(D21) & (D22) & (D23)
\vspace{1mm}
\\
\includegraphics[height=4.5cm]{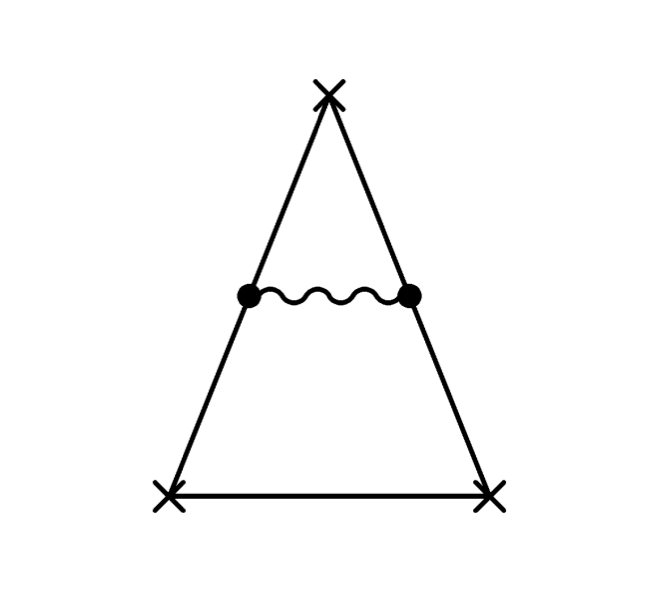} &
\includegraphics[height=4.5cm]{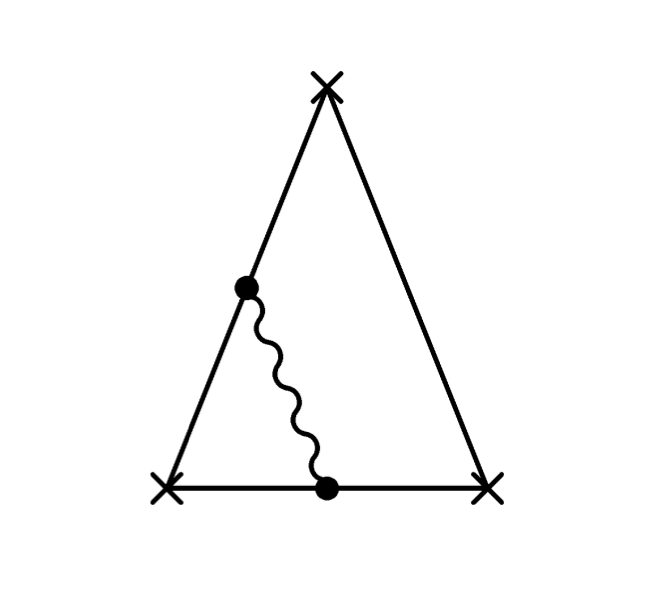} &
\includegraphics[height=4.5cm]{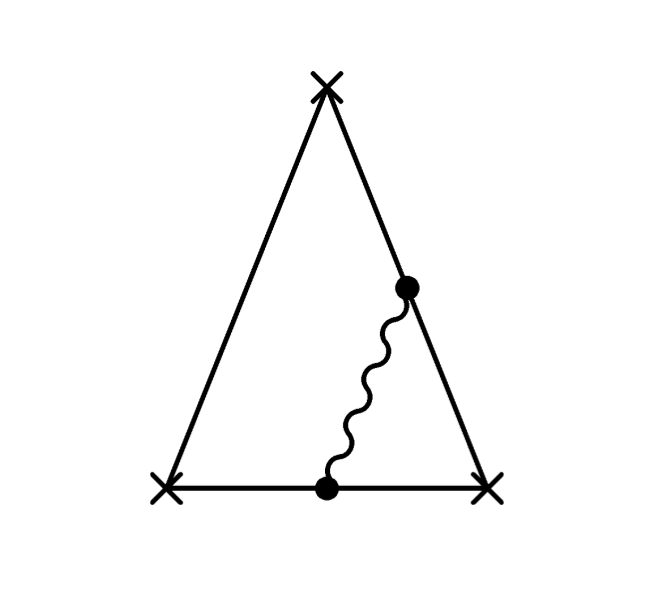}
\vspace{-5mm}
\\
(D31) & (D32) & (D33)
\end{tabular}
\end{center}

Their contributions to the parity-violating tensor structure \eqref{J2J1J1struct} at order $\lambda$, for null polarizations in the limit $x_2 \to x_1$, are given by
\begin{align}
  \frac{\lambda}{\sqrt{N}}\frac{(\pol\cdot x_{12})^2(\pol\cdot x_{13})\e_{\alpha\beta\gamma}x_{12}^{\alpha}x_{13}^{\beta}\pol^{\gamma}}{|x_{13}|^6|x_{12}|^6}
\end{align}
times the following factors,
\begin{align}
  (\mathrm{D}21) &\to \frac{i}{12\pi^4} \ec  &
  (\mathrm{D}22) &\to 0 \ec &
  (\mathrm{D}23) &\to 0 \ec \\
  (\mathrm{D}31) &\to \frac{i}{24\pi^4} \ec &
  (\mathrm{D}32) &\to \frac{-i}{12\pi^4} \ec &
  (\mathrm{D}33) &\to 0 \ed
\end{align}

\section{Conformal Transformations}
\label{conf}

The conformal algebra in Euclidean space is
\begin{align}
  [M_{\mu\nu},P_\rho] &=
    -i(\delta_{\mu\rho}P_\nu - \delta_{\nu\rho}P_\mu) \ec &
  [M_{\mu\nu},K_\rho] &=
    -i(\delta_{\mu\rho}K_\nu - \delta_{\nu\rho}K_\mu) \ec \nonumber\\
  [D,P_\mu] &= -iP_\mu \ec &
  [D,K_\mu] &= iK_\mu \ec \\
  [D,M_{\mu\nu}] &= 0 \ec &
  [K_\mu,P_\nu] &= 2i(\delta_{\mu\nu}D + M_{\mu\nu}) \nonumber\ec
\end{align}
\vspace{-0.85cm}
\begin{align}
  [M_{\mu\nu},M_{\rho\sigma}] &=
    -i \delta_{\mu\rho} M_{\nu\sigma}
    +i \delta_{\nu\rho} M_{\mu\sigma}
    +i \delta_{\mu\sigma} M_{\nu\rho}
    -i \delta_{\nu\sigma} M_{\mu\rho} \ed
\end{align}
The action of $D$ on a local primary operator $\cO(x)$ with dimension $\Delta$ is
\begin{align}
  [D,\cO(0)] &= -i \Delta \cO(0) \,.
\end{align}
The Lorentz generators in the vector representation are
\begin{align}
  ({\tilde M}_{\mu\nu})_{\alpha\beta} &=
    i ( \delta_{\mu\alpha} \delta_{\nu\beta}
        - \delta_{\mu\beta} \delta_{\nu\alpha} )
        \ec
\end{align}
and their action on a tensor operator $J_{\rho_1 \cdots \rho_n}$ is
\begin{align}
  [M_{\mu\nu}, J_{\rho_1 \cdots \rho_n}] &=
  -({\tilde M}_{\mu\nu})_{\rho_1 \alpha} J_{\alpha \rho_2 \cdots \rho_n}
  - \cdots
  -({\tilde M}_{\mu\nu})_{\rho_n \alpha} J_{\rho_1 \rho_2 \cdots \alpha}
  \ed
\end{align}

\section{Anomalous Dimension of $\phi^i\phi^i$}
\label{adim}

 In this appendix we verify explicitly that $J_0 = \phi^i \phi^i / \sqrt{N}$ does not receive an anomalous dimension at two loops and infinite $N$, in accordance with the general results of sections \ref{allorders} and \ref{HScurrents}. To compute the anomalous dimension of $J_0$ we consider the correlator
\begin{align}
  \langle \phi^2(x)\phi^{i_1}(x_1)\phi^{i_2}(x_2)\rangle_{\mathrm{amp.}}
  \label{eq:ADcorr}
\end{align}
in momentum space. The following two diagrams contribute to the divergence:

\vspace{2mm}

\begin{center}
\begin{tabular}{cc}
\includegraphics[width=3.0cm]{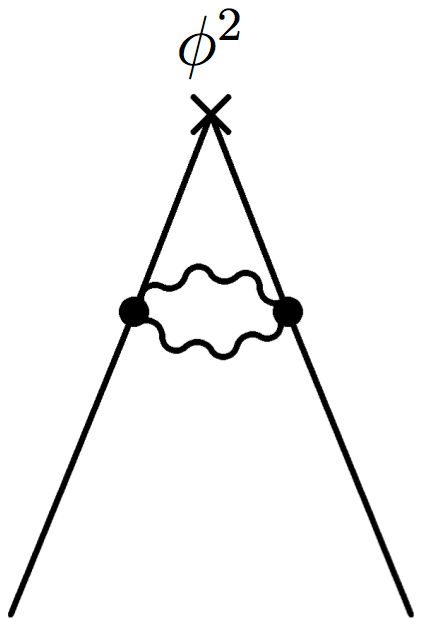} &
\hspace{8mm}
\includegraphics[width=3.0cm]{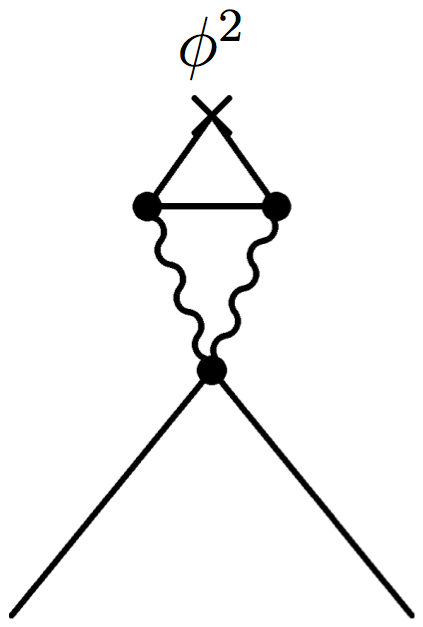}
\\
(C1) &
\hspace{8mm}
(C2)
\end{tabular}
\end{center}

The divergent parts of these diagrams appear in Appendix \ref{diags}. Only (C1) contains a planar diagram, and its contribution at large $N$ to the correlator with amputated $\phi$ legs is
\begin{align}
  \lambda^2 \delta_{i_1i_2} \frac{1}{64\pi^2} \frac{1}{\epsilon}\ed
  \label{adcon}
\end{align}
The bare amputated correlator is related to the amputated correlator of the physical theory by
\begin{gather}
  \left< \phi^2 \phi^{i_1} \phi^{i_2} \right>_{\text{phys.}} =
  \frac{Z_\phi}{Z_{\phi^2}}
  \left< \phi^2 \phi^{i_1} \phi^{i_2} \right>_{\text{bare}} \ec
  \label{corrrel}
\end{gather}
where $J_0^{\text{bare}} = Z_{\phi^2} J_0^{\text{phys.}}$, and $Z_{\phi}=1-\lambda^2\frac{1}{128} \frac{1}{\epsilon}+O(1/N)$ (see \eqref{eq:Zphi}). This should not have any divergence. Using \eqref{adcon}, and noting that the amputated correlator \eqref{eq:ADcorr} equals ($2 \delta_{i_1 i_2}$) at tree-level, the divergence of \eqref{corrrel} at the large $N$ limit, can be seen to be
\begin{align}
  -2 \delta_{i_1i_2} \delta Z_{\phi^2} + O(\lambda^3) \ed
\end{align}
Therefore, to leading order in $1/N$, $\delta Z_{\phi^2} = O(\lambda^3)$  and $\phi^2$ has no anomalous dimension.

For the theory of a complex scalar field in the fundamental representation of $U(N)$, using the relations \eqref{eq:cScalarRen} and the fact that the tree-level correlator equals 1, we also reach the conclusion that the operator $|\phi|^2$ has no anomalous dimension.


\newpage
\begin{thebibliography}{99}

\bibitem{Maldacena:1997re}
  J.~M.~Maldacena,
  ``The Large N limit of superconformal field theories and supergravity,''
  Adv.\ Theor.\ Math.\ Phys.\  {\bf 2 } (1998)  231-252.
  [\href{http://arxiv.org/abs/hep-th/9711200}{hep-th/9711200}].

\bibitem{Sundborg:2000wp}
  B.~Sundborg,
  ``Stringy gravity, interacting tensionless strings and massless higher spins,''
  Nucl.\ Phys.\ Proc.\ Suppl.\  {\bf 102 } (2001)  113-119.
  [\href{http://arxiv.org/abs/hep-th/0103247}{hep-th/0103247}].

\bibitem{Sezgin:2002rt}
  E.~Sezgin, P.~Sundell,
  ``Massless higher spins and holography,''
  Nucl.\ Phys.\  {\bf B644 } (2002)  303-370.
  [\href{http://arxiv.org/abs/hep-th/0205131}{hep-th/0205131}].

\bibitem{Klebanov:2002ja}
  I.~R.~Klebanov, A.~M.~Polyakov,
  ``AdS dual of the critical O(N) vector model,''
  Phys.\ Lett.\  {\bf B550 } (2002)  213-219.
  [\href{http://arxiv.org/abs/hep-th/0210114}{hep-th/0210114}].

\bibitem{Fradkin:1987ks}
  E.~S.~Fradkin, M.~A.~Vasiliev,
  ``On the Gravitational Interaction of Massless Higher Spin Fields,''
  Phys.\ Lett.\  {\bf B189 } (1987)  89-95.


\bibitem{Vasiliev:1999ba}
  M.~A.~Vasiliev,
  ``Higher spin gauge theories: Star product and AdS space,''
  In *Shifman, M.A. (ed.): The many faces of the superworld* 533-610.
  [\href{http://arxiv.org/abs/hep-th/9910096}{hep-th/9910096}].

\bibitem{Giombi:2009wh}
  S.~Giombi, X.~Yin,
  ``Higher Spin Gauge Theory and Holography: The Three-Point Functions,''
  JHEP {\bf 1009 } (2010)  115.
  [\href{http://arxiv.org/abs/0912.3462}{arXiv:0912.3462} [hep-th]].

\bibitem{Giombi:2010vg}
  S.~Giombi, X.~Yin,
  ``Higher Spins in AdS and Twistorial Holography,''
  JHEP {\bf 1104 } (2011)  086.
  [\href{http://arxiv.org/abs/1004.3736}{arXiv:1004.3736} [hep-th]].

\bibitem{Giombi:2011ya}
  S.~Giombi, X.~Yin,
  ``On Higher Spin Gauge Theory and the Critical O(N) Model,''
  [\href{http://arxiv.org/abs/1105.4011}{arXiv:1105.4011} [hep-th]].

\bibitem{Das:2003vw}
  S.~R.~Das, A.~Jevicki,
  ``Large N collective fields and holography,''
  Phys.\ Rev.\  {\bf D68 } (2003)  044011.
  [\href{http://arxiv.org/abs/hep-th/0304093}{hep-th/0304093}].

\bibitem{Koch:2010cy}
  R.~d.~M.~Koch, A.~Jevicki, K.~Jin, J.~P.~Rodrigues,
  ``$AdS_4/CFT_3$ Construction from Collective Fields,''
  Phys.\ Rev.\  {\bf D83 } (2011)  025006.
  [\href{http://arxiv.org/abs/1008.0633}{arXiv:1008.0633} [hep-th]].

\bibitem{Douglas:2010rc}
  M.~R.~Douglas, L.~Mazzucato, S.~S.~Razamat,
  ``Holographic dual of free field theory,''
  Phys.\ Rev.\  {\bf D83 } (2011)  071701.
  [\href{http://arxiv.org/abs/1011.4926}{arXiv:1011.4926} [hep-th]].

\bibitem{Jevicki:2011ss}
  A.~Jevicki, K.~Jin, Q.~Ye,
  ``Collective Dipole Model of AdS/CFT and Higher Spin Gravity,''
  [\href{http://arxiv.org/abs/1106.3983}{arXiv:1106.3983} [hep-th]].

\bibitem{Chen:1992ee}
  W.~Chen, G.~W.~Semenoff, Y.~-S.~Wu,
  ``Two loop analysis of nonAbelian Chern-Simons theory,''
  Phys.\ Rev.\  {\bf D46 } (1992)  5521-5539.
  [\href{http://arxiv.org/abs/hep-th/9209005}{hep-th/9209005}].

\bibitem{Avdeev:1992jt}
  L.~V.~Avdeev, D.~I.~Kazakov, I.~N.~Kondrashuk,
  ``Renormalizations in supersymmetric and nonsupersymmetric nonAbelian Chern-Simons field theories with matter,''
  Nucl.\ Phys.\  {\bf B391 } (1993)  333-357.

\bibitem{Deser:1981wh}
  S.~Deser, R.~Jackiw, S.~Templeton,
  ``Topologically Massive Gauge Theories,''
  Annals Phys.\  {\bf 140}, 372-411 (1982).

\bibitem{Deser:1982vy}
  S.~Deser, R.~Jackiw, S.~Templeton,
  ``Three-Dimensional Massive Gauge Theories,''
  Phys.\ Rev.\ Lett.\  {\bf 48}, 975-978 (1982).

\bibitem{Siegel:1979wq}
  W.~Siegel,
  ``Supersymmetric Dimensional Regularization via Dimensional Reduction,''
  Phys.\ Lett.\  {\bf B84 } (1979)  193.

\bibitem{Gubser:2002vv}
  S.~S.~Gubser, I.~R.~Klebanov,
  ``A Universal result on central charges in the presence of double trace deformations,''
  Nucl.\ Phys.\  {\bf B656 } (2003)  23-36.
  [\href{http://arxiv.org/abs/hep-th/0212138}{hep-th/0212138}].

\bibitem{Petkou:2003zz}
  A.~C.~Petkou,
  ``Evaluating the AdS dual of the critical O(N) vector model,''
  JHEP {\bf 0303 } (2003)  049.
  [\href{http://arxiv.org/abs/hep-th/0302063}{hep-th/0302063}].

\bibitem{Alves:1999hw}
  V.~S.~Alves, M.~Gomes, S.~L.~V.~Pinheiro, A.~J.~da Silva,
  ``A Renormalization group study of the (phi* phi)**3 model coupled to a Chern-Simons field,''
  Phys.\ Rev.\  {\bf D61 } (2000)  065003.
  [\href{http://arxiv.org/abs/hep-th/0001221}{hep-th/0001221}].

\bibitem{deAlbuquerque:2000ec}
  L.~C.~de Albuquerque, M.~Gomes, A.~J.~da Silva,
  ``Renormalization group study of Chern-Simons field coupled to scalar matter in a modified BPHZ subtraction scheme,''
  Phys.\ Rev.\  {\bf D62 } (2000)  085005.
  [\href{http://arxiv.org/abs/hep-th/0005258}{hep-th/0005258}].

\bibitem{Weinberg:1996kr}
  S.~Weinberg,
  ``The quantum theory of fields. Vol. 2: Modern applications,''
  Cambridge, UK: Univ. Pr. (1996) 489 p.


\bibitem{Bardeen:1983rv}
  W.~A.~Bardeen, M.~Moshe, M.~Bander,
  ``Spontaneous Breaking of Scale Invariance and the Ultraviolet Fixed Point in O($n$) Symmetric $(phi^{6}$ in Three-Dimensions) Theory,''
  Phys.\ Rev.\ Lett.\  {\bf 52 } (1984)  1188.

\bibitem{Moshe:2003xn}
  M.~Moshe, J.~Zinn-Justin,
  ``Quantum field theory in the large N limit: A Review,''
  Phys.\ Rept.\  {\bf 385 } (2003)  69-228.
  [\href{http://arxiv.org/abs/hep-th/0306133}{hep-th/0306133}].

\bibitem{Schwarz:2004yj}
  J.~H.~Schwarz,
  ``Superconformal Chern-Simons theories,''
  JHEP {\bf 0411 } (2004)  078.
  [\href{http://arxiv.org/abs/hep-th/0411077}{hep-th/0411077}].

\bibitem{Gaiotto:2007qi}
  D.~Gaiotto, X.~Yin,
  ``Notes on superconformal Chern-Simons-Matter theories,''
  JHEP {\bf 0708 } (2007)  056.
  [\href{http://arxiv.org/abs/0704.3740}{arXiv:0704.3740} [hep-th]].

\bibitem{Amit:1984ri}
  D.~J.~Amit, E.~Rabinovici,
  ``BREAKING OF SCALE INVARIANCE IN phi**6 THEORY: TRICRITICALITY AND CRITICAL END POINTS,''
  Nucl.\ Phys.\  {\bf B257}, 371 (1985).

\bibitem{Dias:2003pw}
  A.~G.~Dias, M.~Gomes, A.~J.~da Silva,
  ``Dynamical breakdown of symmetry in (2+1) dimensional model containing the Chern-Simons field,''
  Phys.\ Rev.\  {\bf D69 } (2004)  065011.
  [\href{http://arxiv.org/abs/hep-th/0305043}{hep-th/0305043}].

\bibitem{Dias:2010it}
  A.~G.~Dias, A.~F.~Ferrari,
  ``Renormalization Group and Conformal Symmetry Breaking in the Chern-Simons Theory Coupled to Matter,''
  Phys.\ Rev.\  {\bf D82 } (2010)  085006.
  [\href{http://arxiv.org/abs/1006.5672}{arXiv:1006.5672} [hep-th]].

\bibitem{Rabinovici:2011jj}
  E.~Rabinovici, M.~Smolkin,
  ``On the dynamical generation of the Maxwell term and scale invariance,''
  JHEP {\bf 1107}, 040 (2011).
  [\href{http://arXiv.org/abs/arXiv:1102.5035}{arXiv:1102.5035} [hep-th]].

\bibitem{Ferrari:2010ex}
  A.~F.~Ferrari, E.~A.~Gallegos, M.~Gomes, A.~C.~Lehum, J.~R.~Nascimento, A.~Y.~.Petrov, A.~J.~da Silva,
  ``Coleman-Weinberg mechanism in a three-dimensional supersymmetric Chern-Simons-Matter model,''
  Phys.\ Rev.\  {\bf D82}, 025002 (2010).
  [\href{http://arXiv.org/abs/arXiv:1004.0982}{arXiv:1004.0982} [hep-th]].

\bibitem{Girardello:2002pp}
  L.~Girardello, M.~Porrati, A.~Zaffaroni,
  ``3-D interacting CFTs and generalized Higgs phenomenon in higher spin theories on AdS,''
  Phys.\ Lett.\  {\bf B561 } (2003)  289-293.
  [\href{http://arxiv.org/abs/hep-th/0212181}{hep-th/0212181}].

\bibitem{Heidenreich:1980xi}
  W.~Heidenreich,
  ``Tensor Products Of Positive Energy Representations Of So(3,2) And So(4,2),''
  J.\ Math.\ Phys.\  {\bf 22 } (1981)  1566.

\bibitem{Elitzur:2005kz}
  S.~Elitzur, A.~Giveon, M.~Porrati, E.~Rabinovici,
  ``Multitrace deformations of vector and adjoint theories and their holographic duals,''
  JHEP {\bf 0602 } (2006)  006.
  [\href{http://arxiv.org/abs/hep-th/0511061}{hep-th/0511061}].

\bibitem{Witten:2001ua}
  E.~Witten,
  ``Multitrace operators, boundary conditions, and AdS / CFT correspondence,''
  [\href{http://arxiv.org/abs/hep-th/0112258}{hep-th/0112258}].

\bibitem{Berkooz:2002ug}
  M.~Berkooz, A.~Sever, A.~Shomer,
  ``'Double trace' deformations, boundary conditions and space-time singularities,''
  JHEP {\bf 0205 } (2002)  034.
  [\href{http://arxiv.org/abs/hep-th/0112264}{hep-th/0112264}].

\bibitem{Giombi:2011kc}
  S.~Giombi, S.~Minwalla, S.~Prakash, S.~P.~Trivedi, S.~R.~Wadia, X.~Yin,
  ``Chern-Simons Theory with Vector Fermion Matter,''
  [\href{http://arXiv.org/abs/arXiv:1110.4386}{arXiv:1110.4386} [hep-th]].

\bibitem{Giombi:2011rz}
  S.~Giombi, S.~Prakash, X.~Yin,
  ``A Note on CFT Correlators in Three Dimensions,''
  [\href{http://arxiv.org/abs/1104.4317}{arXiv:1104.4317} [hep-th]].

\bibitem{Boos:1987bg}
  E.~E.~Boos, A.~I.~Davydychev,
  ``A Method Of The Evaluation Of The Vertex Type Feynman Integrals,''
  Moscow Univ.\ Phys.\ Bull.\  {\bf 42N3 } (1987)  6-10.

\bibitem{Shenker:2011zf}
  S.~H.~Shenker, X.~Yin,
  ``Vector Models in the Singlet Sector at Finite Temperature,''
  [\href{http://arxiv.org/abs/1109.3519}{arXiv:1109.3519} [hep-th]].

\bibitem{Gopakumar:1998ki}
  R.~Gopakumar, C.~Vafa,
  ``On the gauge theory / geometry correspondence,''
  Adv.\ Theor.\ Math.\ Phys.\  {\bf 3 } (1999)  1415-1443.
  [\href{http://arxiv.org/abs/hep-th/9811131}{hep-th/9811131}].

\bibitem{Sinha:2000ap}
  S.~Sinha, C.~Vafa,
  ``SO and Sp Chern-Simons at large N,''
  [\href{http://arxiv.org/abs/hep-th/0012136}{hep-th/0012136}].

\bibitem{Leigh:2003gk}
  R.~G.~Leigh, A.~C.~Petkou,
  ``Holography of the N=1 higher spin theory on AdS(4),''
  JHEP {\bf 0306 } (2003)  011.
  [\href{http://arxiv.org/abs/hep-th/0304217}{hep-th/0304217}].

\bibitem{Sezgin:2003pt}
  E.~Sezgin, P.~Sundell,
  ``Holography in 4D (super) higher spin theories and a test via cubic scalar couplings,''
  JHEP {\bf 0507 } (2005)  044.
  [\href{http://arxiv.org/abs/hep-th/0305040}{hep-th/0305040}].

\end{thebibliography}
\end{document}